%% file: _main.tex
\documentclass{egpubl}
\usepackage{eurovis2023}

\SpecialIssuePaper         

\usepackage[T1]{fontenc}
\usepackage{dfadobe}  

\usepackage{cite}  
\BibtexOrBiblatex
\electronicVersion
\PrintedOrElectronic
\ifpdf \usepackage[pdftex]{graphicx} \pdfcompresslevel=9
\else \usepackage[dvips]{graphicx} \fi

\usepackage{egweblnk}

\usepackage{multirow}
\usepackage{tabu}                      
\usepackage{booktabs}                  
\usepackage[table]{xcolor}
\usepackage{xspace}
\usepackage{dblfloatfix} 
\usepackage{svg}
\usepackage{tabularx}

\title[Visualization Corpora for Automated Chart Analysis]%
      {The State of the Art in Creating Visualization Corpora\\for Automated Chart Analysis}

\author[Chen Chen \& Zhicheng Liu]
{\parbox{\textwidth}{\centering Chen Chen
\orcid{0000-0003-3171-0657}
        and Zhicheng Liu
        \orcid{0000-0002-1015-2759} 
        }
        \\
{\parbox{\textwidth}{\centering 
Department of Computer Science, University of Maryland College Park
\\
       }
}
}

\newcommand{\bpstart}[1]{\noindent{\textbf{#1}}}

\newcommand{\eg}{{e.g.,}\xspace}
\newcommand{\ie}{{i.e.,}\xspace}
\newcommand{\etal}{{et al.}\xspace}

\usepackage{url}

\usepackage{breakurl}

\usepackage{adjustbox}
\usepackage{array}
\newcolumntype{R}[2]{%
    >{\adjustbox{angle=#1,lap=\width-(#2)}\bgroup}%
    l%
    <{\egroup}%
}
\newcommand*\rot{\multicolumn{1}{R{60}{1em}}}

\definecolor{thisRed}{RGB}{213, 94, 0}
\definecolor{thisRed2}{RGB}{230, 159, 0}
\definecolor{thisBlue}{RGB}{0, 114, 178}
\definecolor{thisBlue2}{RGB}{86,180,233}

\newcommand{\taskd}[1]{\textcolor{thisBlue}{#1}}
\newcommand{\task}[1]{\textcolor{thisBlue2}{#1}}
\newcommand{\property}[1]{\textcolor{thisRed}{#1}}
\newcommand{\propertyd}[1]{\textcolor{thisRed2}{#1}}

\newcommand{\revise}[1]{\textcolor{black}{#1}}

\RequirePackage{fontawesome}

\begin{document}


\maketitle
\begin{abstract}
   We present a state-of-the-art report on visualization corpora in automated chart analysis research. We survey 56 papers that created or used a visualization corpus as the input of their research techniques or systems. 
   Based on a multi-level task taxonomy that identifies the goal, method, and outputs of automated chart analysis, 
   we examine the property space of existing chart corpora along five dimensions: format, scope, collection method, annotations, and diversity. Through the survey, we summarize common patterns and practices of creating chart corpora, identify research gaps and opportunities, and discuss the desired properties of future benchmark corpora and the required tools to create them.


\end{abstract}  

\input{sections/1_intro.tex}

\input{sections/2_method.tex}

\input{sections/3_task.tex}

\input{sections/4_format.tex}

\input{sections/5_scope.tex}
\input{sections/6_collection.tex}

\input{sections/7_annotation.tex}

\input{sections/8_diversity.tex}

\input{sections/9_challenges_and_opportunities.tex}

\input{sections/10_conclusion.tex}

\bibliographystyle{eg-alpha-doi}
\bibliography{egbibsample}


\newpage
\input{sections/ShortBio}


\end{document}

%% file: sections/1_intro.tex
\section{Introduction}\label{sec:1}

Recent advances in automated chart analysis techniques~\cite{li2022structure,cui2021mixed,battle2018beagle,dibia2019data2vis,kahou2017figureqa,poco2017reverse} seek to enable more effective retrieval, interpretation, creation, and transformation of data visualizations. Typically, these research efforts require a corpus of charts collected from the wild. Such corpora are essential for developing and evaluating chart analysis methods, and for providing real-world examples that end users can modify and repurpose.

There has been, however, little research on 1) the common practices for creating the corpora, 2) what constitutes a good chart corpus for various tasks and applications, and 3) the potential pitfalls and gaps in existing corpus-based research for automated chart analysis. 
Based on our preliminary observation, many relevant papers do not use corpora contributed by prior work; instead, they build their own corpora. There are many possible reasons for this: previous corpora are not publicly available~\cite{davila2020chart}, the corpora are not of high quality~\cite{lai2020automatic}, the corpora do not have the labels required for specific tasks, or the existing corpora do not contain visualizations representing the desired design space. The current state of corpora creation and usage seems \textit{ad hoc}, making 
it difficult to compare chart analysis techniques, measure scientific progress, and identify unsolved research problems. 


This survey aims to contribute a comprehensive understanding of the state of the art in creating corpora for automated chart analysis research. 
By ``chart'' we refer to two-dimensional statistical data graphics or infographics without 3D effects
We collect 56 research papers from areas including AI, HCI, NLP, and Visualization that either contribute a new chart corpus, or a technique or system that takes charts in a corpus as inputs, or a model trained on a corpus. We first identify the automated chart analysis tasks along three dimensions: \textit{why} (the goal), \textit{how} (the method), and \textit{what} (the outputs). We then extract five main properties of chart corpora used in these research works: \textit{chart format}, \textit{corpus scope}, \textit{collection method}, \textit{annotations}, and \textit{diversity}. Along these task dimensions and corpus properties, we present results on the current patterns and practices of corpora creation and usage.
Through the survey, we identify research gaps and opportunities in corpus-based automated chart analysis, recommend desired properties of new corpora to be created to support the research investigations, and discuss research ideas on tools and methods for creating the desired benchmark corpora.

\vspace{-0.3cm}

\subsection{Related Surveys}
To the best of our knowledge, there is no comprehensive survey on the corpora used in automated chart analysis. Two surveys, AI4VIS~\cite{wu2021ai4vis} and ML4VIS~\cite{wang2021survey}, have reviewed current literature on AI-empowered and ML-based approaches for data visualization, respectively. However, both focus on categorizing tasks or techniques and do not discuss the impacts of corpora. Also, most works included in the two surveys are based on machine learning and neural networks; thus other heuristics-based approaches may be missing there.

The most relevant discussions were found in Deng \etal~\cite{deng2022visimages} and Davila \etal~\cite{davila2020chart}. The former provides a general review of existing visualization corpora to motivate their goal of creating a new one, and the latter describes the different levels of automation observed in chart corpora creation; an in-depth analysis of corpus properties, however, is not included. This STAR will fill the gap by describing the property space of chart corpora in detail, identifying standard practices of creating corpora, establishing guidelines for the curation process, and suggesting research problems that can benefit from high-quality reusable corpora.




%% file: sections/2_method.tex
\section{Survey Methods}

In this section, we describe the search criteria and process, our coding process, and the analysis method.

\subsection{Search Criteria and Process}

We first started with the papers included in two recent surveys on artificial intelligence approaches~\cite{wu2021ai4vis} and machine learning methods~\cite{wang2021survey} for data visualization. Both surveys cover publications from a variety of disciplines, such as Visualization, Human-Computer Interaction~(HCI), Artificial Intelligence~(AI), and Natural Language Processing~(NLP). We chose these two surveys as our starting point to collect relevant papers that contribute or adopt chart corpora because (1) many works covered in these two surveys introduce techniques or systems to create, analyze, or reason about charts, thus requiring visualization corpora; and (2) the methods or models presented in these papers include classic ML techniques~(such as random forest~\cite{biau2016random} and support vector machine~\cite{cortes1995support}), modern neural networks~(including graph neural networks~\cite{wu2020comprehensive}), and heuristics-based algorithms, which impose varying requirements on the desired corpora. Starting from these two collections allows us to form an initial set of diverse chart corpora. 
Specifically, we set three criteria to filter papers in the two repositories based on the scope of this survey defined in the introduction:
\begin{enumerate}
    \item The primary contribution of the paper is either a chart corpus~(e.g., Jobin~\etal~\cite{jobin2019docfigure} and Kahou~\etal\cite{kahou2017figureqa}), or a technique or system that takes the collected charts as inputs~(e.g., Savva~\etal~\cite{savva2011revision}), or a model trained on the collected charts (e.g., Cui~\etal~\cite{cui2019text}). We thus did not include papers like VizNet~\cite{hu2019viznet} and the work by Haehn \etal \cite{haehn2018evaluating} because the former introduced a corpus of data tables, and the latter presented an empirical study instead of a system or technique.
    \item The corpus is described explicitly in the paper or supplementary materials. We used keywords including \textit{bitmap, svg, dataset, corpus, training, crawl, search} to search for descriptions for a corpus within each paper. This criterion allows us to obtain first-hand accurate information from the authors about their corpora, the original descriptions of which would serve as the foundation of the subsequent coding and analysis processes.
    \item The corpus consists of 2-dimensional static charts or infographics. We exclude 3D visualizations such as scientific visualizations because generating and analyzing such visualizations lead to very different research problems and outputs~\cite{wu2021ai4vis,xu2020survey}. Corpora containing scientific equations~\cite{lee2017viziometrics},  color ramps~\cite{smart2019color} and hand-drawn sketches~\cite{ma2020ladv} were also excluded. 
\end{enumerate}

\begin{figure}[ht]
\centering
\includegraphics[width=0.475\textwidth]{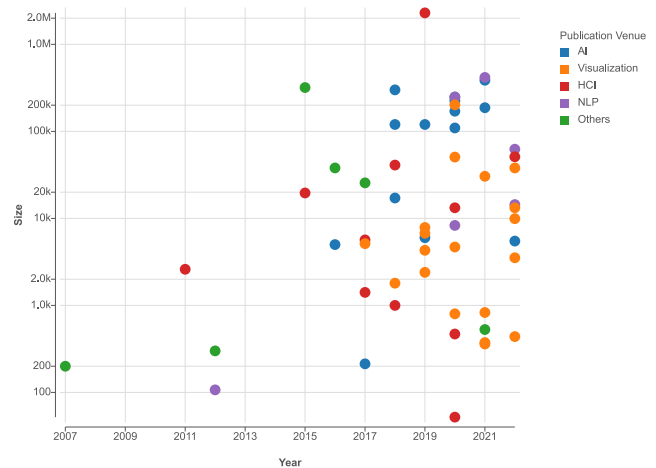}
\caption{An overview of 56 chart corpora we collected. Each dot in the visualization represents a chart corpus, whose x-axis represents the publication year, y-axis represents the corpus size (in log scale), and color represents the publication area. Out of the 56 chart corpora, 19 are from Visualization (\eg IEEE VIS, TVCG), 14 are from AI (\eg AAAI, CVPR, IJCNN) , 11 are from HCI (\eg ACM CHI, UIST), 6 are from NLP (\eg ACL, EMNLP) , and 6 are from other areas (\eg WWW, ICIP, ECML-PKDD).}
\label{fig:corporaOverview}
\end{figure}

Following these three criteria, we obtained an initial set of 41 papers that introduce visualization corpora. We then applied one round of relation-search approach~\cite{mcnabb2017survey} (i.e., graph traversal over the citation and reference networks~\cite{hu2019vizml}), to augment the initial paper set. During this process, the above three criteria were still enforced. 
To focus on the latest development in chart corpora and be consistent with the initial paper set's year range, we did not include papers published before 2007. This procedure added 15 more papers, resulting in a final set of 56 chart corpora. In Figure~\ref{fig:corporaOverview}, we show an overview of collected 56 corpora in terms of corpus size, publication venue, and year. It can be seen that recent corpora tend to have large sizes, and the three most frequent publication venues are Visualization, AI, and HCI.

We acknowledge that although the search criteria are clearly defined, our manual search has limitations. It is possible that some related papers are not included. However, unlike most state-of-the-art survey articles that discuss and analyze visualization techniques, our goal is to summarize current practices in creating corpora for automated chart analysis. Thus, instead of exhaustively finding all the qualified visualization corpora, a sufficiently diverse sample can allow us to perform a comprehensive analysis.

\subsection{Coding and Analysis}
Our analysis starts with investigating why and how the corpora are used in these papers. Specifically, we identify the \textit{research tasks} presented in each paper, which dictate the curation of the corpus (\eg
what chart types and visual styles to collect and what kinds of labels or annotations are needed). For example, 
the primary goal of Revision~\cite{savva2011revision} was to 
 perform chart type classification and chart redesign; it thus collected 10 types of single-view visualizations and required labels on chart type and text element position.  The model or technique used in a paper also influences aspects of corpus curation such as the format of the input charts. 
For example, Li~\etal~\cite{li2022structure} proposed a two-thread neural network model which takes both the bitmap and SVG representations of a chart, and Data2Vis~\cite{dibia2019data2vis} used a sequence-to-sequence recurrent neural network model which takes a chart in the Vega-Lite specification representation. To this end, we follow the well-established what-why-how dimensions to categorize tasks introduced in the collected papers across three levels:

\bpstart{Why: the goal.} This task level describes the purpose and applications of automated chart analysis. 

\bpstart{How: the method.} This task level describes the techniques or mechanisms to analyze a chart.
Oftentimes multiple techniques are used in conjunction to achieve a goal. 

\bpstart{What: the output.} This task level describes the outputs of chart analysis methods. They can be at a holistic level (\eg chart type), or at a finer granularity (\eg elements such as axis and encodings).

We further identify five key properties of visualization corpora that are frequently included in the authors' descriptions and are most important to the above three-level tasks:
\begin{itemize}
    \item \bpstart{Format} refers to the file type of charts in a corpus, which includes bitmap graphics~(.jpeg, .png), vector graphics~(.svg), and programs (e.g., Vega-Lite specification). 
    \item \bpstart{Scope} refers to the selection criteria and assumptions about the charts in a corpus. These are usually specified to constrain the research problem space. 
    \item \bpstart{Collection method} specifies how a corpus was collected, which is influenced by both \textit{Format} and \textit{Scope}.
    \item \bpstart{Annotations} are labels associated with charts, serving as ground truth for automated chart analysis tasks.
    \item \bpstart{Diversity} measures how much the charts differ from one another within a corpus.
\end{itemize}

For each task level and each corpus property, we adopted a bottom-up coding approach.
One author first performed the following two coding jobs: (1) categorizing and labeling the task levels and property dimensions, during which the corresponding taxonomies were built iteratively, and (2) recording chart corpus descriptions from surveyed papers for each corpus property. Labels for new-coming papers or corpora were verified to see if they fit into existing categories, and if not, both authors discussed together to verify again and establish an alternative task if needed.
After this first round of paper and corpora coding, both authors examined the coding results; whenever conflicts of understanding arose, the two authors proceeded to discuss the cases until reaching a consensus for every paper and corpus.
Our supplemental materials contain the details of our analysis, with quotes from the papers to demonstrate the validity of the coding.

\revise{
Our final coding results are presented in Table~\ref{tbl:overview}, where the rows represent surveyed papers~(corpora) grouped by their task goals, and the columns represent  methods, outputs, and corpora properties. Section~\ref{sec:task} includes the descriptions for the fine-grained categorizations of the task taxonomy and the corpus proprieties. The size and public link~(if any) of each chart corpus are also included.
}

\input{figures/table1.tex}

%% file: figures/table1.tex
\begin{table*}[!ht]
    \centering
    \aboverulesep=0ex 
    \belowrulesep=0ex 
    \caption{All the surveyed corpora organized by research goal, along with their methods, outputs, and properties. Some descriptive properties, such as scope of design variations, annotation types, and diversity, are not included.}
    \resizebox{.95\textwidth}{!}{
    \begin{tabular}{c|c|c|c|c|c|c|c|r|c|c|c|c|c|c|c|c|c|c|c|c|c}
        
         \multicolumn{1}{c}{}
        & \rot{\taskd{Modern Neural Network}}
        & \rot{\taskd{Classic machine learning}}
        & \rot{\taskd{Heuristics-based Algorithm}}
        & \rot{\task{Chart Component}}
        & \rot{\task{Synthesized Description}}
        & \rot{\task{Derived Property}}
        & \multicolumn{1}{c}{}
        & \multicolumn{1}{c}{}
        & \multicolumn{1}{c}{}
        & \rot{\property{Number of Chart Types} }
        & \rot{\propertyd{Vector Graphics}}
        & \rot{\propertyd{Bitmap Graphics}}
        & \rot{\propertyd{Visualization Program}}
        & \rot{\property{Reusing and Transforming}}
        & \rot{\property{Web Crawling}}
        & \rot{\property{Manual Curation}} 
        & \rot{\property{Computer-Aided Generation}}
        & \rot{\propertyd{In-house Labeling}}
        & \rot{\propertyd{Crowdsourcing}}
        & \rot{\propertyd{Template-based Generation}} 
        & \rot{\propertyd{Automatic Extraction} }
        \\
        \toprule
        \multicolumn{1}{c}{\textbf{\task{Goal}}}
        & 
        \multicolumn{3}{c|}{\textbf{\taskd{Method}}}
        & 
        \multicolumn{3}{c|}{\textbf{\task{Output}}}
        &
        \multicolumn{1}{c|}{\textbf{Corpus}}
        & 
        \multicolumn{1}{c|}{\textbf{Size}}
        & 
        \multicolumn{1}{c|}{\textbf{Link}}
        & 
        \multicolumn{1}{c|}{\textbf{\property{Scope}}}
        &  
        \multicolumn{3}{c|}{\textbf{\propertyd{Format}}}
        & 
        \multicolumn{4}{c|}{\textbf{\property{Collection Method}}}
        & 
        \multicolumn{4}{c}{\textbf{\propertyd{Annotation Method}}}
        \\
        \toprule
         \multirow{6}{*}[-1em]{\task{\shortstack{Create a \\ chart  corpus}}}
         & 
         & \cellcolor{thisBlue}
         & 
         & \cellcolor{thisBlue2}
         & 
         & 
         & \cite{battle2018beagle}
         & 41K
         & \href{https://homes.cs.washington.edu/~leibatt/beagle.html}{\faExternalLink}
         & \property{24}
         & \cellcolor{thisRed2}
         & 
         & 
         & 
         & \cellcolor{thisRed}
         & 
         & 
         & \cellcolor{thisRed2}
         & 
         &
         &
         \\
         \cmidrule{2-22}
         & \cellcolor{thisBlue}
         & 
         & 
         & 
         & \cellcolor{thisBlue2}
         & 
         & \cite{masry2022chartqa}
         & 14K
         & \href{https://github.com/vis-nlp/ChartQA}{\faExternalLink}
         & \property{3}
         & \cellcolor{thisRed2}
         & \cellcolor{thisRed2}
         & 
         & 
         & \cellcolor{thisRed}
         & 
         & 
         & 
         & \cellcolor{thisRed2}
         &
         &
         \\
         \cmidrule{2-22}
         & \cellcolor{thisBlue}
         & 
         & 
         & 
         & \cellcolor{thisBlue2}
         & 
         & \cite{kahou2017figureqa}
         & 120K
         & \href{https://github.com/Maluuba/FigureQA}{\faExternalLink}
         & \property{3}
         & 
         & \cellcolor{thisRed2}
         & 
         & 
         & 
         & 
         & \cellcolor{thisRed}
         & 
         & 
         & \cellcolor{thisRed2}
         &
         \\
         \cmidrule{2-22}
         & \cellcolor{thisBlue}
         & 
         & 
         & 
         & \cellcolor{thisBlue2}
         & 
         & \cite{mathew2022infographicvqa}
         & 5.5K
         & \href{https://www.docvqa.org/publications}{\faExternalLink}
         & \property{1}
         & 
         & \cellcolor{thisRed2}
         & 
         & 
         & 
         & \cellcolor{thisRed}
         & 
         & \cellcolor{thisRed2}
         & 
         &
         &
         \\
         \cmidrule{2-22}
         & \cellcolor{thisBlue}
         & 
         & 
         & 
         & \cellcolor{thisBlue2}
         & 
         & \cite{mahinpei2022linecap}
         & 3.5K
         & \href{https://github.com/anita76/LineCapDataset}{\faExternalLink}
         & \property{1}
         & 
         & \cellcolor{thisRed2}
         & 
         & \cellcolor{thisRed}
         & 
         & 
         & 
         & 
         & \cellcolor{thisRed2}
         &
         &
         \\
         \cmidrule{2-22}
         & \cellcolor{thisBlue}
         & 
         & \cellcolor{thisBlue}
         & 
         & \cellcolor{thisBlue2}
         & 
         & \cite{chang2022mapqa}
         & 62K
         & \href{https://github.com/OSU-slatelab/MapQA}{\faExternalLink}
         & \property{1}
         & 
         & \cellcolor{thisRed2}
         & 
         & 
         & \cellcolor{thisRed}
         & 
         & \cellcolor{thisRed}
         & 
         & 
         & \cellcolor{thisRed2}
         &
         \\
         \cmidrule{2-22}
         & \cellcolor{thisBlue}
         & 
         & 
         & \cellcolor{thisBlue2}
         & 
         & 
         & \cite{deng2022visimages}
         & 38K
         & \href{https://visimages.github.io/}{\faExternalLink}
         & \property{34}
         & 
         & \cellcolor{thisRed2}
         & 
         & 
         & 
         & \cellcolor{thisRed}
         & 
         & \cellcolor{thisRed2}
         & 
         &
         &
         \\
         \toprule
         \multirow{4}{*}[-0.2em]{\task{\shortstack{Generate \\ chart designs \\automatically}}}
         & \cellcolor{thisBlue}
         & 
         & 
         & 
         & 
         & \cellcolor{thisBlue2}
         & \cite{zhao2020chartseer}
         & 10K
         & 
         & \property{6}
         & 
         & 
         & \cellcolor{thisRed2}
         & \cellcolor{thisRed}
         & 
         & 
         & \cellcolor{thisRed}
         & 
         & 
         &
         &
         \\
         \cmidrule{2-22}
         & \cellcolor{thisBlue}
         & 
         & 
         & 
         & 
         & \cellcolor{thisBlue2}
         & \cite{dibia2019data2vis}
         & 4.3K
         & 
         & \property{5}
         & 
         & 
         & \cellcolor{thisRed2}
         & \cellcolor{thisRed}
         & 
         & 
         & 
         & 
         & 
         &
         &
         \\
         \cmidrule{2-22}
         & \cellcolor{thisBlue}
         & 
         & \cellcolor{thisBlue}
         & \cellcolor{thisBlue2}
         & 
         & 
         & \cite{cui2019text}
         & 800
         & 
         & \property{1}
         & 
         & \cellcolor{thisRed2}
         & 
         & 
         & 
         & \cellcolor{thisRed}
         & 
         & 
         & 
         &
         &
         \\
         \cmidrule{2-22}
         & \cellcolor{thisBlue}
         & 
         & 
         & \cellcolor{thisBlue2}
         & 
         & \cellcolor{thisBlue2}
         & \cite{hu2019vizml}
         & 2.3M
         & \href{https://vizml.media.mit.edu/}{\faExternalLink}
         & \property{3}
         & 
         & \cellcolor{thisRed2}
         & 
         & 
         & \cellcolor{thisRed}
         & 
         & 
         & 
         & 
         &
         &
         \\
         \toprule
         \multirow{5}{*}[-0.2em]{\task{\shortstack{Retrieve charts \\ matching \\ certain criteria}}}
         & 
         & 
         & \cellcolor{thisBlue}
         & \cellcolor{thisBlue2}
         & 
         & 
         & \cite{chen2020composition}
         & 360
         & 
         & \property{--}
         & 
         & \cellcolor{thisRed2}
         & 
         & 
         & \cellcolor{thisRed}
         & 
         & 
         & \cellcolor{thisRed2}
         & 
         &
         &
         \\
         \cmidrule{2-22}
         & 
         & 
         & \cellcolor{thisBlue}
         & \cellcolor{thisBlue2}
         & 
         & 
         & \cite{chen2015diagramflyer}
         & 319K
         & 
         & \property{3}
         & 
         & \cellcolor{thisRed2}
         & 
         & 
         & \cellcolor{thisRed}
         & 
         & 
         & 
         & 
         &
         &
         \\
         \cmidrule{2-22}
         & 
         & 
         & \cellcolor{thisBlue}
         & \cellcolor{thisBlue2}
         & 
         & 
         & \cite{hoque2019searching}
         & 7.9K
         & 
         & \property{--}
         & \cellcolor{thisRed2}
         & 
         & 
         & 
         & \cellcolor{thisRed}
         & 
         & 
         & 
         & 
         &
         &
         \\
         \cmidrule{2-22}
         & \cellcolor{thisBlue}
         & 
         & 
         & 
         & 
         & \cellcolor{thisBlue2}
         & \cite{li2022structure}
         & 51K
         & 
         & \property{5}
         & \cellcolor{thisRed2}
         & \cellcolor{thisRed2}
         & 
         & 
         & \cellcolor{thisRed}
         & 
         & 
         & 
         & 
         &
         &
         \\
         \cmidrule{2-22}
         & \cellcolor{thisBlue}
         & \cellcolor{thisBlue}
         & 
         & 
         & 
         & \cellcolor{thisBlue2}
         & \cite{oppermann2020vizcommender}
         & 4.7K
         & 
         & \property{--}
         & 
         & 
         & \cellcolor{thisRed2}
         & 
         & \cellcolor{thisRed}
         & 
         & 
         & 
         & 
         &
         &
         \\
         \toprule
         \multirow{8}{*}[-0.5em]{\task{\shortstack{Modify an \\ existing chart}}}
         & 
         & 
         & \cellcolor{thisBlue}
         & \cellcolor{thisBlue2}
         & 
         & 
         & \cite{cui2021mixed}
         & 438
         & 
         & \property{1}
         & \cellcolor{thisRed2}
         & 
         & 
         & \cellcolor{thisRed}
         & 
         & \cellcolor{thisRed}
         & 
         & 
         & 
         &
         &
         \\
         \cmidrule{2-22}
         & 
         & 
         & \cellcolor{thisBlue}
         & \cellcolor{thisBlue2}
         & 
         & 
         & \cite{poco2017extracting}
         & 1.8K
         & 
         & \property{7}
         & 
         & \cellcolor{thisRed2}
         & 
         & 
         & \cellcolor{thisRed}
         & 
         & 
         & \cellcolor{thisRed2}
         & 
         &
         &
         \\
         \cmidrule{2-22}
         & 
         & 
         & \cellcolor{thisBlue}
         & \cellcolor{thisBlue2}
         & 
         & 
         & \cite{yuan2021infocolorizer}
         & 13K
         & 
         & \property{1}
         & 
         & \cellcolor{thisRed2}
         & 
         & \cellcolor{thisRed}
         & 
         & 
         & 
         & 
         & 
         &
         &
         \\
         \cmidrule{2-22}
         & 
         & 
         & \cellcolor{thisBlue}
         & \cellcolor{thisBlue2}
         & 
         & 
         & \cite{wu2020mobilevisfixer}
         & 374
         & 
         & \property{--}
         & \cellcolor{thisRed2}
         & 
         & 
         & 
         & \cellcolor{thisRed}
         & 
         & 
         & 
         & 
         &
         &
         \\
         \cmidrule{2-22}
         & 
         & 
         & \cellcolor{thisBlue}
         & \cellcolor{thisBlue2}
         & 
         & 
         & \cite{qian2020retrieve}
         & 829
         & 
         & \property{1}
         & 
         & \cellcolor{thisRed2}
         & 
         & 
         & 
         & \cellcolor{thisRed}
         & 
         & \cellcolor{thisRed2}
         & 
         &
         &
         \\
         \cmidrule{2-22}
         & 
         & \cellcolor{thisBlue}
         & \cellcolor{thisBlue}
         & \cellcolor{thisBlue2}
         & 
         & 
         & \cite{savva2011revision}
         & 2.6K
         & \href{http://idl.cs.washington.edu/files/revision-corpus.zip}{\faExternalLink}
         & \property{10}
         & 
         & \cellcolor{thisRed2}
         & 
         & 
         & \cellcolor{thisRed}
         & 
         & 
         & \cellcolor{thisRed2}
         & 
         &
         &
         \\
         \cmidrule{2-22}
         & 
         & 
         & \cellcolor{thisBlue}
         & \cellcolor{thisBlue2}
         & 
         & 
         & \cite{chen2019towards}
         & 4.7K
         & 
         & \property{1}
         & 
         & \cellcolor{thisRed2}
         & 
         & 
         & \cellcolor{thisRed}
         & 
         & \cellcolor{thisRed}
         & \cellcolor{thisRed2}
         & 
         &
         &
         \\
         \cmidrule{2-22}
         & \cellcolor{thisBlue}
         & 
         & \cellcolor{thisBlue}
         & \cellcolor{thisBlue2}
         & 
         & 
         & \cite{huang2021visual}
         & 187K
         & 
         & \property{1}
         & 
         & \cellcolor{thisRed2}
         & 
         & 
         & \cellcolor{thisRed}
         & 
         & 
         & 
         & 
         & 
         & \cellcolor{thisRed2}
         \\
         \toprule
         \multirow{11}{*}[-0.4em]{\task{\shortstack{Generate \\natural language \\descriptions}}}
         & 
         & 
         & \cellcolor{thisBlue}
         & \cellcolor{thisBlue2}
         & \cellcolor{thisBlue2}
         & 
         & \cite{kim2020answering}
         & 52
         & 
         & \property{2}
         & \cellcolor{thisRed2}
         & \cellcolor{thisRed2}
         & \cellcolor{thisRed2}
         & 
         & 
         & \cellcolor{thisRed}
         & 
         & 
         & \cellcolor{thisRed2}
         &
         &
         \\
         \cmidrule{2-22}
         & \cellcolor{thisBlue}
         & 
         & 
         & 
         & \cellcolor{thisBlue2}
         & 
         & \cite{sharma2019chartnet}
         & 6K
         & 
         & \property{2}
         & 
         & \cellcolor{thisRed2}
         & 
         & 
         & 
         & 
         & \cellcolor{thisRed}
         & \cellcolor{thisRed2}
         & 
         &
         &
         \\
         \cmidrule{2-22}
         & \cellcolor{thisBlue}
         & 
         & 
         & 
         & \cellcolor{thisBlue2}
         & 
         & \cite{obeid2020chart}
         & 8.3K
         & \href{https://github.com/JasonObeid/Chart2Text}{\faExternalLink}
         & \property{2}
         & 
         & \cellcolor{thisRed2}
         & 
         & 
         & \cellcolor{thisRed}
         & 
         & 
         & \cellcolor{thisRed2}
         & 
         &
         &
         \\
         \cmidrule{2-22}
         & \cellcolor{thisBlue}
         & 
         & 
         & 
         & \cellcolor{thisBlue2}
         & 
         & \cite{kafle2018dvqa}
         & 300K
         & \href{https://github.com/kushalkafle/DVQA_dataset}{\faExternalLink}
         & \property{1}
         & 
         & \cellcolor{thisRed2}
         & 
         & 
         & 
         & 
         & \cellcolor{thisRed}
         & 
         & 
         & \cellcolor{thisRed2}
         &
         \\
         \cmidrule{2-22}
         & \cellcolor{thisBlue}
         & 
         & 
         & 
         & \cellcolor{thisBlue2}
         & 
         & \cite{chen2019figure}
         & 110K
         & 
         & \property{3}
         & 
         & \cellcolor{thisRed2}
         & 
         & \cellcolor{thisRed}
         & 
         & 
         & 
         & 
         & 
         & \cellcolor{thisRed2}
         &
         \\
         \cmidrule{2-22}
         & \cellcolor{thisBlue}
         & 
         & 
         & 
         & \cellcolor{thisBlue2}
         & 
         & \cite{reddy2019figurenet}
         & 120K
         & \href{https://github.com/Maluuba/FigureQA}{\faExternalLink}
         & \property{3}
         & 
         & \cellcolor{thisRed2}
         & 
         & \cellcolor{thisRed}
         & 
         & 
         & 
         & 
         & 
         &
         &
         \\
         \cmidrule{2-22}
         & \cellcolor{thisBlue}
         & 
         & 
         & 
         & \cellcolor{thisBlue2}
         & 
         & \cite{chaudhry2020leaf}
         & 248K
         & 
         & \property{6}
         & 
         & \cellcolor{thisRed2}
         & 
         & 
         & 
         & 
         & \cellcolor{thisRed}
         & 
         & 
         & \cellcolor{thisRed2}
         &
         \\
         \cmidrule{2-22}
         & \cellcolor{thisBlue}
         & 
         & \cellcolor{thisBlue}
         & \cellcolor{thisBlue2}
         & \cellcolor{thisBlue2}
         & 
         & \cite{methani2020plotqa}
         & 224K
         & \href{https://github.com/NiteshMethani/PlotQA}{\faExternalLink}
         & \property{3}
         & 
         & \cellcolor{thisRed2}
         & 
         & 
         & 
         & 
         & \cellcolor{thisRed}
         & 
         & \cellcolor{thisRed2}
         & 
         & \cellcolor{thisRed2}
         \\
         \cmidrule{2-22}
         & \cellcolor{thisBlue}
         & 
         & 
         & 
         & \cellcolor{thisBlue2}
         & 
         & \cite{hsu2021scicap}
         & 417K
         & 
         & \property{1}
         & 
         & \cellcolor{thisRed2}
         & 
         & \cellcolor{thisRed}
         & 
         & \cellcolor{thisRed}
         & 
         & 
         & 
         &
         &
         \\
         \cmidrule{2-22}
         & \cellcolor{thisBlue}
         & 
         & 
         & 
         & \cellcolor{thisBlue2}
         & 
         & \cite{singh2020stl}
         & 248K
         & \href{https://github.com/Maluuba/FigureQA}{\faExternalLink}
         & \property{1}
         & 
         & \cellcolor{thisRed2}
         & 
         & \cellcolor{thisRed}
         & 
         & 
         & 
         & 
         & 
         &
         &
         \\
         \cmidrule{2-22}
         & 
         & 
         & \cellcolor{thisBlue}
         & \cellcolor{thisBlue2}
         & \cellcolor{thisBlue2}
         & 
         & \cite{demir2012summarizing}
         & 107
         & 
         & \property{1}
         & \cellcolor{thisRed2}
         & 
         & 
         & 
         & 
         & \cellcolor{thisRed}
         & 
         & 
         & 
         &
         &
         \\
         \toprule
         \multirow{11}{*}[-6em]{\task{\shortstack{Extract \\chart semantics}}}
         & 
         & \cellcolor{thisBlue}
         & 
         & \cellcolor{thisBlue2}
         & 
         & 
         & \cite{al2017machine}
         & 213
         & 
         & \property{1}
         & 
         & \cellcolor{thisRed2}
         & 
         & 
         & 
         & \cellcolor{thisRed}
         & 
         & 
         & 
         &
         &
         \\
         \cmidrule{2-22}
         & 
         & \cellcolor{thisBlue}
         & \cellcolor{thisBlue}
         & \cellcolor{thisBlue2}
         & 
         & 
         & \cite{huang2007system}
         & 200
         & 
         & \property{3}
         & 
         & \cellcolor{thisRed2}
         & 
         & 
         & 
         & \cellcolor{thisRed}
         & 
         & 
         & 
         &
         &
         \\
         \cmidrule{2-22}
         & \cellcolor{thisBlue}
         & 
         & \cellcolor{thisBlue}
         & \cellcolor{thisBlue2}
         & 
         & 
         & \cite{lai2020automatic}
         & 469
         & 
         & \property{3}
         & 
         & \cellcolor{thisRed2}
         & 
         & 
         & 
         & \cellcolor{thisRed}
         & 
         & \cellcolor{thisRed2}
         & 
         &
         &
         \\
         \cmidrule{2-22}
         & \cellcolor{thisBlue}
         & 
         & \cellcolor{thisBlue}
         & \cellcolor{thisBlue2}
         & 
         & 
         & \cite{luo2021chartocr}
         & 387K
         & \href{https://github.com/soap117/DeepRule}{\faExternalLink}
         & \property{3}
         & 
         & \cellcolor{thisRed2}
         & 
         & 
         & \cellcolor{thisRed}
         & 
         & 
         & 
         & 
         & 
         & \cellcolor{thisRed2}
         \\
         \cmidrule{2-22}
         & \cellcolor{thisBlue}
         & 
         & \cellcolor{thisBlue}
         & \cellcolor{thisBlue2}
         & 
         & 
         & \cite{rane2021chartreader}
         & 528
         & \href{https://drive.google.com/drive/u/0/folders/1M8kwdQE7bpjpdT08ldBURFdzLaQR9n5h}{\faExternalLink}
         & \property{1}
         & 
         & \cellcolor{thisRed2}
         & 
         & 
         & 
         & \cellcolor{thisRed}
         & 
         & 
         & 
         &
         &
         \\
         \cmidrule{2-22}
         & \cellcolor{thisBlue}
         & 
         & \cellcolor{thisBlue}
         & \cellcolor{thisBlue2}
         & 
         & 
         & \cite{jung2017chartsense}
         & 5.7K
         & 
         & \property{10}
         & \cellcolor{thisRed2}
         & 
         & \cellcolor{thisRed}
         & \cellcolor{thisRed}
         & 
         & 
         & 
         & 
         &
         &
         &
         \\
         \cmidrule{2-22}
         & \cellcolor{thisBlue}
         & 
         & 
         & \cellcolor{thisBlue2}
         & 
         & \cellcolor{thisBlue2}
         & \cite{tang2016deepchart}
         & 5K
         & 
         & \property{5}
         & 
         & \cellcolor{thisRed2}
         & 
         & 
         & 
         & \cellcolor{thisRed}
         & 
         & 
         & 
         &
         &
         \\
         \cmidrule{2-22}
         & \cellcolor{thisBlue}
         & \cellcolor{thisBlue}
         & 
         & \cellcolor{thisBlue2}
         & 
         & 
         & \cite{chagas2018evaluation}
         & 17K
         & 
         & \property{10}
         & 
         & \cellcolor{thisRed2}
         & 
         & 
         & 
         & \cellcolor{thisRed}
         & \cellcolor{thisRed}
         & 
         & 
         &
         &
         \\
         \cmidrule{2-22}
         & \cellcolor{thisBlue}
         & 
         & \cellcolor{thisBlue}
         & \cellcolor{thisBlue2}
         & 
         & 
         & \cite{lu2020exploring}
         & 13K
         & 
         & \property{1}
         & 
         & \cellcolor{thisRed2}
         & 
         & 
         & 
         & \cellcolor{thisRed}
         & 
         & \cellcolor{thisRed2}
         & 
         &
         &
         \\
         \cmidrule{2-22}
         & \cellcolor{thisBlue}
         & 
         & 
         & \cellcolor{thisBlue2}
         & 
         & 
         & \cite{luo2021hybrid}
         & 170K
         & 
         & \property{1}
         & 
         & \cellcolor{thisRed2}
         & 
         & 
         & \cellcolor{thisRed}
         & 
         & 
         & 
         & 
         &
         & \cellcolor{thisRed2}
         \\
         \cmidrule{2-22}
         & 
         & \cellcolor{thisBlue}
         & 
         & 
         & 
         & \cellcolor{thisBlue2}
         & \cite{saleh2015learning}
         & 20K
         & 
         & \property{1}
         & 
         & \cellcolor{thisRed2}
         & 
         & 
         & 
         & \cellcolor{thisRed}
         & 
         & 
         & \cellcolor{thisRed2}
         &
         &
         \\
         \cmidrule{2-22}
         & \cellcolor{thisBlue}
         & 
         & 
         & 
         & 
         & \cellcolor{thisBlue2}
         & \cite{bylinskii2017learning}
         & 1.4K
         & 
         & \property{3}
         & 
         & \cellcolor{thisRed2}
         & 
         & \cellcolor{thisRed}
         & 
         & 
         & 
         & 
         & \cellcolor{thisRed2}
         &
         &
         \\
         \cmidrule{2-22}
         & \cellcolor{thisBlue}
         & 
         & 
         & \cellcolor{thisBlue2}
         & 
         & 
         & \cite{kim2018multimodal}
         & 1K
         & 
         & \property{1}
         & 
         & \cellcolor{thisRed2}
         & 
         & \cellcolor{thisRed}
         & 
         & 
         & 
         & 
         & \cellcolor{thisRed2}
         &
         &
         \\
         \cmidrule{2-22}
         & \cellcolor{thisBlue}
         & 
         & 
         & \cellcolor{thisBlue2}
         & 
         & 
         & \cite{zhou2021reverse}
         & 3K
         & 
         & \property{1}
         & 
         & \cellcolor{thisRed2}
         & 
         & 
         & 
         & \cellcolor{thisRed}
         & \cellcolor{thisRed}
         & \cellcolor{thisRed2}
         & 
         &
         & \cellcolor{thisRed2}
         \\
         \cmidrule{2-22}
         & \cellcolor{thisBlue}
         & \cellcolor{thisBlue}
         & \cellcolor{thisBlue}
         & \cellcolor{thisBlue2}
         & \cellcolor{thisBlue2}
         & 
         & \cite{poco2017reverse}
         & 5K
         & \href{https://github.com/uwdata/rev}{\faExternalLink}
         & \property{4}
         & \cellcolor{thisRed2}
         & \cellcolor{thisRed2}
         & \cellcolor{thisRed2}
         & 
         & \cellcolor{thisRed}
         & 
         & \cellcolor{thisRed}
         & \cellcolor{thisRed2}
         & 
         &
         &
         \\
         \cmidrule{2-22}
         & 
         & \cellcolor{thisBlue}
         & \cellcolor{thisBlue}
         & \cellcolor{thisBlue2}
         & 
         & 
         & \cite{choudhury2016scalable}
         & 3.8K
         & 
         & \property{2}
         & \cellcolor{thisRed2}
         & \cellcolor{thisRed2}
         & 
         & 
         & \cellcolor{thisRed}
         & 
         & 
         & \cellcolor{thisRed2}
         & 
         &
         &
         \\
         \cmidrule{2-22}
         & \cellcolor{thisBlue}
         & 
         & \cellcolor{thisBlue}
         & \cellcolor{thisBlue2}
         & 
         & 
         & \cite{cliche2017scatteract}
         & 2.7K
         & 
         & \property{1}
         & 
         & \cellcolor{thisRed2}
         & 
         & 
         & 
         & 
         & \cellcolor{thisRed}
         & 
         & 
         &
         & \cellcolor{thisRed2}
         \\
         \cmidrule{2-22}
         & \cellcolor{thisBlue}
         & 
         & 
         & 
         & 
         & \cellcolor{thisBlue2}
         & \cite{ma2018scatternet}
         & 51K
         & 
         & \property{1}
         & 
         & \cellcolor{thisRed2}
         & 
         & 
         & 
         & 
         & \cellcolor{thisRed}
         & \cellcolor{thisRed2}
         & 
         &
         &
         \\
         \cmidrule{2-22}
         & 
         & \cellcolor{thisBlue}
         & \cellcolor{thisBlue}
         & \cellcolor{thisBlue2}
         & 
         & 
         & \cite{gao2012view}
         & 300
         & 
         & \property{3}
         & 
         & \cellcolor{thisRed2}
         & 
         & 
         & 
         & \cellcolor{thisRed}
         & 
         & 
         & 
         &
         &
         \\
         \cmidrule{2-22}
         & \cellcolor{thisBlue}
         & \cellcolor{thisBlue}
         & 
         & 
         & 
         & \cellcolor{thisBlue2}
         & \cite{fu2019visualization}
         & 6.7K
         & 
         & \property{1}
         & 
         & \cellcolor{thisRed2}
         & 
         & \cellcolor{thisRed}
         & 
         & 
         & 
         & \cellcolor{thisRed2}
         & 
         &
         &
         \\
         \cmidrule{2-22}
         & \cellcolor{thisBlue}
         & \cellcolor{thisBlue}
         & \cellcolor{thisBlue}
         & \cellcolor{thisBlue2}
         & 
         & 
         & \cite{choi2019visualizing}
         & 2.4K
         & 
         & \property{3}
         & 
         & \cellcolor{thisRed2}
         & 
         & 
         & \cellcolor{thisRed}
         & 
         & 
         & 
         & 
         &
         &
         \\
        \bottomrule
    \end{tabular}}
    \label{tbl:overview}
\end{table*}

%% file: sections/3_task.tex
\section{Tasks: Why, How, and What}\label{sec:task}
\bpstart{Why: the goal.}
We went through the collected papers, unified their vocabularies about their research goals and tasks, and identified 6 categories for the goal dimension:
\begin{itemize}
    \item \textit{Create a chart corpus}, which aims at introducing a benchmark chart collection for certain chart types or analysis tasks. For example, the Beagle corpus~\cite{battle2018beagle} consists of SVG visualizations collected from five popular charting tools on the web; the LineCap corpus~\cite{mahinpei2022linecap} curated line charts with figure captioning; and the MapQA corpus~\cite{chang2022mapqa} introduced a question answering annotations specifically for choropleth maps.
    \item \textit{Extract chart semantics}, where ``semantics'' refers to information spanning a range of concepts, 
    including
    low-level primitives like the mark type~\cite{luo2021hybrid} and attributes~\cite{poco2017reverse}, the role of a mark group (e.g., axis, glyph), and
    high-level meta-information such as chart type~\cite{jung2017chartsense} or the underlying dataset~\cite{ma2018scatternet}. 
    The extracted semantics are useful for various 
    downstream applications.
    \item \textit{Modify an existing chart}, which transforms a chart for new contexts or needs. 
    There are two kinds of modification: (1) \textit{reusing} a chart where the underlying data is modified but the visual designs are maintained~\cite{chen2019towards,cui2021mixed,qian2020retrieve}, and (2) \textit{redesigning} a chart where the visual mappings and styles are changed while the underlying data is untouched~\cite{poco2017extracting,wu2020mobilevisfixer,savva2011revision}. Modifying a chart usually requires explicit extraction of certain chart semantics. 
    \item \textit{Generate chart designs automatically}, which learns from a corpus of charts about existing visual mappings and automatically generates new charts for a given dataset or task ~\cite{dibia2019data2vis,hu2019vizml,zhao2020chartseer,cui2019text}.
    \item \textit{Retrieve charts matching certain criteria}, which is about searching from a chart database for charts that (1) share some common characteristics (\eg visual styles, structures, or topics) with a reference chart~\cite{li2022structure,zhao2020chartseer}, or (2) are semantically related to some keyword queries~\cite{chen2015diagramflyer,hoque2019searching}.
    \item \textit{Generate natural language descriptions}, where information in the charts is transformed into natural language forms for purposes such as enhanced accessibility - in many contexts, it is easier to consume and ask questions about a chart when the information is presented in a verbal or audio format~\cite{reddy2019figurenet,chen2019figure, obeid2020chart,demir2012summarizing, kahou2017figureqa,chaudhry2020leaf}. 
    
\end{itemize}

\bpstart{How: the method.}
We have observed 3 categories of methods: 
\begin{itemize}
    \item \textit{Modern neural networks~(NN)}, including convolutional neural networks~\cite{gu2018recent} used by~Chagas~\etal~\cite{chagas2018evaluation} and Cui~\etal~\cite{cui2019text}, recurrent neural networks~\cite{sutskever2014sequence} used by~Dibia and Demiralp~\cite{dibia2019data2vis} and Zhao~\etal~\cite{zhao2020chartseer}, and graph neural networks~\cite{wu2020comprehensive} used by~Li~\etal\cite{li2022structure}.
    \item \textit{Classic machine learning~(ML) models}.  For example, to label collected chart images automatically, Battle~\etal~\cite{battle2018beagle} adopted random forest~\cite{biau2016random} to perform automatic chart type classification; Poco and Heer~\cite{poco2017reverse} used support vector machines~(SVM)~\cite{cortes1995support} to classify texts presented in a chart into their roles like axis label and legend title.
    \item \textit{Heuristics-based algorithms,} which usually are human-crafted rules designed for specific tasks. For example, Cui~\etal~\cite{cui2021mixed} abstracted six types of chart element update schemes with respect to the underlying data values and developed a chart reusing algorithm; Hoque and Agrawala~\cite{hoque2019searching} introduced a heuristic chart deconstructor to obtain visual styles and structures by utilizing the encoded data values encapsulated in D3 visualizations.
\end{itemize}
These methods can also be used in conjunction to finish multi-stage tasks. 
For example, to understand chart semantics, both Revision~\cite{savva2011revision} and ChartSense~\cite{jung2017chartsense} first used \textit{neural networks} to classify the type of chart, and then applied mark-specific \textit{heuristics-based algorithms} to extract the underlying data.

\bpstart{What: the output.}
We have identified the following outputs of automated chart analysis methods:
\begin{itemize}
    \item \textit{Chart components}
    \begin{itemize}
        \item mark types (\eg bar, circle)~\cite{poco2017reverse, savva2011revision} and roles (\eg whether a rectangle is a bar or a legend)~\cite{al2017machine,rane2021chartreader}
        \item text elements~\cite{luo2021hybrid,al2017machine} and roles (\eg annotation, axis label)~\cite{huang2007system,choudhury2016scalable}
        \item reference marks such as axis \& legend~\cite{poco2017extracting,wu2020mobilevisfixer}
        \item chart type~\cite{battle2018beagle,tang2016deepchart,chagas2018evaluation,poco2017reverse,choudhury2016scalable}
        \item source data~\cite{savva2011revision, jung2017chartsense}
        \item mark grouping (\eg marks belonging to a glyph, grouped bars in a stacked bar chart)~\cite{cui2021mixed,li2022structure}
        \item encodings (\ie the mappings between data fields and visual channels)~\cite{poco2017reverse,harper2017converting}
        \item layouts (\ie how marks or glyphs are arranged spatially)~\cite{chen2019towards,chen2020composition}
    \end{itemize}
    \item \textit{Synthesized descriptions}
    \begin{itemize}
        \item captions~\cite{reddy2019figurenet,chen2019figure}
        \item text summaries~\cite{obeid2020chart,demir2012summarizing}
        \item answers to questions on charts~\cite{kahou2017figureqa,chaudhry2020leaf}
    \end{itemize}
    \item \textit{Derived properties}
    \begin{itemize}
        \item vectorized representation~\cite{li2022structure, zhao2020chartseer}
        \item chart style similarity~\cite{saleh2015learning,ma2018scatternet}
        \item chart topic similarity~\cite{oppermann2020vizcommender}
        \item chart quality~\cite{fu2019visualization}
        \item visual salience of marks and texts~\cite{bylinskii2017learning}
    \end{itemize}
\end{itemize}


Again, these outputs are not exclusive to each other, i.e., one task can output multiple components. For example, Revision and ChartSense output both chart type and source data.

We now take the REV~\cite{poco2017reverse} system as an example to illustrate how a chart corpus is created and used in practical applications based on the task taxonomies identified above. The goal of REV is to \textit{extract chart semantics}. More specifically, it extracts visual encoding specifications for a given chart. To do this, Poco and Heer collected a corpus from three different sources. For each chart in the corpus, they annotated the chart~(mark) type, and the bounding boxes, contents, and roles of the text elements. They then built an end-to-end pipeline to process the charts: OCR-based \textit{heuristics} that output \textit{text localizations and contents}; a multi-class \textit{support vector machine} that outputs \textit{text roles}; a \textit{convolutional neural network} that outputs \textit{mark type classifications}; and finally a set of \textit{heuristics} that outputs \textit{data type}, \textit{domain}, \textit{range}, and \textit{scale type for each axis}. At each stage, the corresponding annotations in the corpus are used to evaluate the model performance and report the statistics.




%% file: sections/4_format.tex
\section{Chart Format}\label{sec:format}

Chart format refers to the file type of charts in a corpus, which includes bitmap graphics~(.jpeg, .png)~\cite{savva2011revision,tang2016deepchart,jung2017chartsense,poco2017reverse,poco2017extracting,deng2022visimages}, vector graphics~(.svg)~\cite{harper2014deconstructing,harper2017converting,battle2018beagle,wu2020mobilevisfixer,li2022structure}, and programs~(code snippets that generate charts)~\cite{dibia2019data2vis,zhao2020chartseer,kim2020answering}. 

\begin{figure}[ht]
\centering
\includegraphics[width=0.475\textwidth]{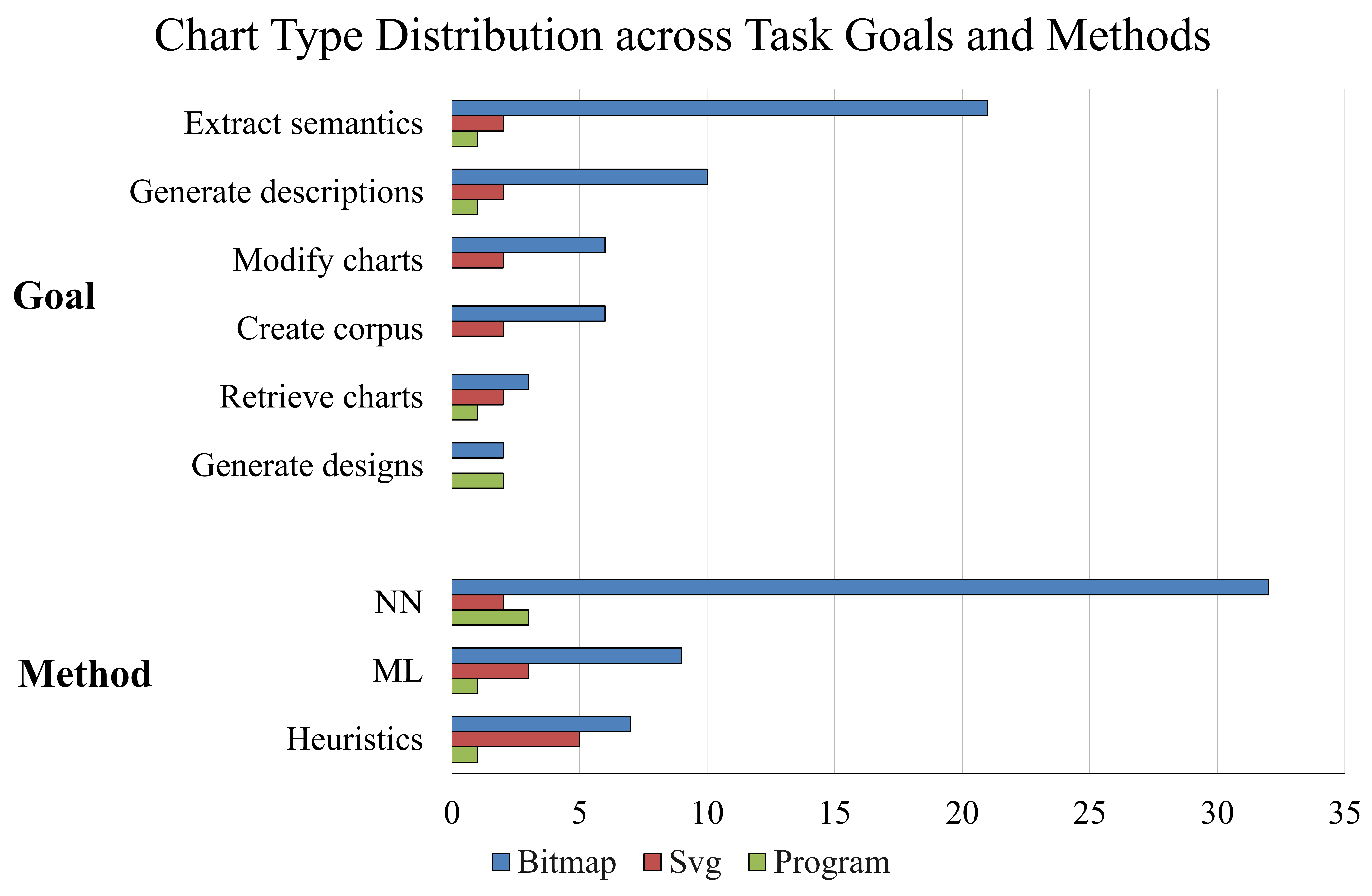}
\caption{The usage frequency of each chart format across different task goals and methods. The y-axis labels are set to abbreviations for task purposes and methods to save space.
}
\label{fig:format2whyANDhow}
\end{figure}

Out of the 56 papers we collected, $48$ used bitmap-graphics corpora, $10$ used vector-graphics corpora, and $5$ used program corpora. Five works
mixed the usage of multiple chart formats: Poco and Heer~\cite{poco2017reverse} and Kim~\etal~\cite{kim2020answering} used charts of all three formats, and Masry~\etal~\cite{masry2022chartqa}, Choudhury~\etal~\cite{choudhury2016scalable}, and Li~\etal~\cite{li2022structure} used both the bitmap and vector graphics. The usage frequency of each type across different task purposes and methods is presented in Figure~\ref{fig:format2whyANDhow}. In cases where multiple methods are used, the primary method performed on a corpus is decided following the order of NN, ML, then Heuristics. 

As the least used format, programs were adopted in five works in our collection, and four out of them used the language Vega~\cite{satyanarayan2015reactive} or Vega-Lite~\cite{satyanarayan2016vega}~(which is in the JSON format): Poco and Heer~\cite{poco2017reverse} analyzed Vega’s scene graph to extract bounding boxes and role labels for texts automatically; from Vega-Lite Specifications, Kim~\etal~\cite{kim2020answering} extracts encodings and underlying data table which are later used to develop question answering techniques; both ChartSeer~\cite{zhao2020chartseer} and Data2Vis~\cite{dibia2019data2vis} apply RNN~\cite{sutskever2014sequence} to Vega-Lite specifications to retrieve next-step visualization recommendations during EDA and candidate charts based on given datasets, respectively. The remaining one, VizCommender~\cite{oppermann2020vizcommender}, measured similarity between pair-wise visualization specifications in online Tableau workbooks to recommend to the user similar visualizations or workbooks.
It can be seen that these works either take advantage of abundant semantic-related information presented in the programs or regard the programs as texts and directly apply learning-based sequential models. Thus, the usage scenario of programs highly depends on their language-specific grammar and structures, limiting their generalizability to a broader range of charts and tasks. The remaining of this section will focus on the comparison between the bitmap and vector formats.



\begin{figure*}[!b]
    \centering
    \includegraphics[width=\textwidth]{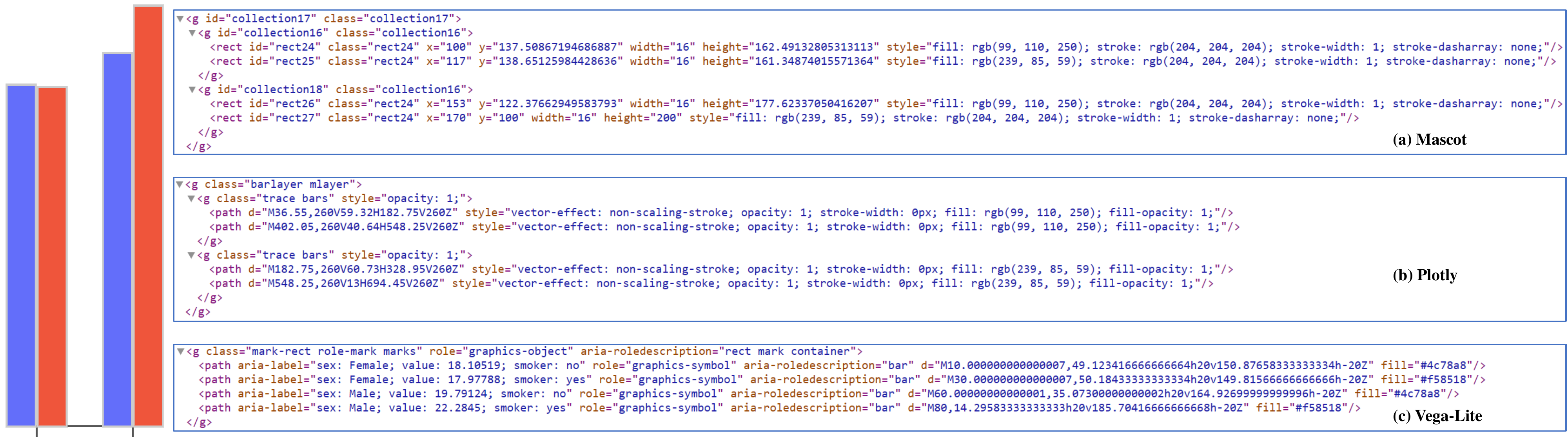}
    \caption{SVG representations of the same grouped bar chart created with (a) Mascot~\cite{liu2021atlas}, (b) Plotly~\cite{Plotly}, and (c) Vega-Lite~\cite{satyanarayan2016vega} using the data from~\cite{GroupBarChartFromPlotly}. To save space, only the $<$g$>$ elements containing rectangle marks are shown.}
    \label{fig:svgInconsistency}
    \end{figure*}

\bpstart{Availability of Chart Semantic Information.} 
The bitmap graphics format only records the pixel information of a chart without providing directly-accessible semantic information ranging from low-level details (\eg marks, roles of visual elements as well as visual encodings) to high-level information like graphical elements grouping, chart type and underlying dataset(s). The vector-based SVG format, on the other hand, is less lossy and embeds certain low-level semantic details in its XML structure, including visual elements types~(\eg text, line, rect, circle, path) and visual styles~(\eg stroke, fill, opacity, x, y, width, height, radius). 

Researchers thus try to leverage the semantic information provided by the vector graphics format whenever available. 
For example, Poco and Heer~\cite{poco2017reverse} collected charts from the news website
Quartz~\cite{Quartz} where the SVG format is available. 
They first extracted texts from the SVG files and then fed the charts into a GUI to obtain annotated bitmap images; in ChartQA~\cite{masry2022chartqa}, whenever the SVG format is available, the authors extracted the bounding boxes of different graphical elements~(e.g., x-axis labels) from the SVG files to train their data extraction models. Without access to SVG-format charts, one can only extract text elements as well as their bounding boxes through human annotations~\cite{qian2020retrieve,deng2022visimages} or OCR-based techniques~\cite{huang2007system,lai2020automatic,luo2021chartocr,masry2022chartqa,poco2017extracting,kim2018multimodal}. In some other works, the technique pipeline can be made more concise if the input images in the vector graphics format are available; e.g., the Revision system~\cite{savva2011revision} has an intermediate mark extraction step that applies heuristics to detect rectangle and circle marks in an input chart, which can be achieved by parsing the elements with \textit{rect} and \textit{circle} tags if its corresponding SVG image is available.

In some special cases, there exists additional high-level information embedded in vector graphics images. For example, SVG images generated by the D3 library~\cite{bostock2011d3} embed the source data using the \textit{\_\_data\_\_} attributes of SVG elements. It is thus possible to directly access the source data and extract semantics more easily using decomposition, restyling, and retrieval algorithms~\cite{harper2014deconstructing,harper2017converting,hoque2019searching}.
These algorithms, however, work exclusively on D3 SVG charts, and cannot be applied to SVG charts generated using other tools or collected from other sources. 

Despite the support to embed semantic information in the SVG format, 
the embedded semantics in vector graphics images from the wild is not always accurate or reliable. We identify three sources of noise and uncertainty:
\begin{itemize}
    \item SVG element tags do not always accurately reflect the semantic mark type, i.e., the same mark type may be represented using different SVG elements. For instance, bar charts in the SVG format created with Mascot.js (previously known as Atlas.js~\cite{liu2021atlas}) are composed of \textit{<rect>} elements, while those created with Vega-Lite library~\cite{satyanarayan2016vega} are using \textit{<path>} elements~(shown in Figure~\ref{fig:svgInconsistency}). Circle marks are also represented differently in these two tools. Thus, preprocessing is needed to detect the semantic mark type.
    \item Inconsistent Grouping of SVG elements is observed across different visualization tools: the $<$g$>$ element is often adopted to group SVG elements in diverse---sometimes random---ways~\cite{choudhury2016scalable}, and the grouping does not necessarily reflect the desired semantics. 
    For example, 
    grouped bar charts created with different tools have different grouping structures in their SVG format.
    In Figure~\ref{fig:svgInconsistency}, three grouped bar chart examples created with Mascot~\cite{liu2021atlas}, Plotly~\cite{Plotly}, and Vega-Lite~\cite{satyanarayan2016vega}, respectively, are presented together with their corresponding SVG files. It can be seen that these three tools groups rectangles in different ways: Mascot groups rectangles sharing the same x-axis label under a $<g>$ element, Plotly groups rectangles sharing the same fill color, and Vega-Lite groups all rectangles together in one $<g>$ element. 
    
    Similarly, groupings for axe and legend elements are unpredictable as well: in some examples, axis labels are put into the same group, while in others, each label forms its own group with its corresponding tick mark. Poco and Heer~\cite{poco2017reverse} also reported that the title in an SVG chart could be composed of several texts specified as separate elements, requiring additional efforts to detect and merge them. Thus, to obtain the real semantic grouping for elements in a given SVG image, one cannot solely rely on the given SVG hierarchical structure; appropriate clustering or classification algorithms are needed.
    \item An SVG scene graph sometimes contains noisy elements: to perform chart analysis tasks, one may need to first distinguish between visualization marks that form the main chart, and graphical objects that are not part of the main chart content like off-screen tooltips used for interaction, transparent background marks, and random watermarks drew as $<path>$ elements.
\end{itemize}

Note that these uncertainties and noises don't exist in bitmap images: different representations and grouping of graphical marks, as well as invisible noisy elements presented in the SVG file will not influence the rendered bitmap image. 

\bpstart{Compatibility with Different Models.} One advantage of bitmap charts is that they are naturally compatible with modern convolutional neural networks~(CNNs). Figure~\ref{fig:format2whyANDhow} shows NN is the most frequently used method on bitmap-based corpora, which indicates that the bitmap graphics format is usually the first choice for many end-to-end neural-network-based systems~\cite{tang2016deepchart,chagas2018evaluation,luo2021hybrid,mathew2022infographicvqa,mahinpei2022linecap,li2022structure,fu2019visualization,huang2021visual,ma2018scatternet,chang2022mapqa,chen2019figure,rane2021chartreader}. In these cases, usually some additional preprocessing steps, such as image cropping \& resizing~\cite{choi2019visualizing} and data augmentation~\cite{kim2018multimodal,zhao2020chartseer}, are needed.
In contrast, SVG charts in their XML structure cannot be directly fed into CNNs.

However, SVG charts have been shown to have potential as inputs for graph neural networks~(GNNs), since SVG elements are organized as trees that can be generalized as graphs. Li~\etal~\cite{li2022structure} performed feature engineering based on the embedded semantics from SVG charts, constructed SVG-based graphs, and ran GNN-based contrastive learning~\cite{sun2019infograph}. The output representations are combined with visual representations obtained from bitmap-based CNN learning to retrieve charts of similar visual appearance as well as structure. This work is a first step towards SVG-based graph learning, which remains an open direction. Besides, as we previously discussed, the availability of chart semantic information allows people to handcraft meaningful features, and then develop classic learning-based methods~\cite{battle2018beagle,choudhury2016scalable} or rule-based heuristics~\cite{cui2021mixed,demir2012summarizing}, to perform chart analysis tasks; this pipeline is more commonly observed in our collected papers than applying NN on SVGs as shown in Figure~\ref{fig:format2whyANDhow}.



\bpstart{Interactivity Support.} 
In addition to the embedded semantic information, the SVG format standard is developed for the web, and is designed to work well with with other web standards such as CSS (Cascading Style Sheets), DOM (Document Object Model), and JavaScript. SVG-based charts can thus be easily enhanced with interactive features, which can be beneficial in the following ways:
\begin{itemize}
    \item \textit{Corpus Creation}: SVG charts make it easy to develop interactive annotation tools where people can collaborate with computers to make the labeling process less laborious~\cite{chen2019towards}.
    \item \textit{Interactive Interfaces}:
    SVG charts as the input make the development of mixed-initiative pipelines~\cite{allen1999mixed} easier; e.g., the Chartreuse system~\cite{cui2021mixed} presents a PowerPoint add-in where users can select chart elements, review chart decomposition results, modify data items, and update the selected chart.
\end{itemize}




%% file: sections/5_scope.tex
\section{Scope: Chart Type and Design Variation}
Scope refers to the assumptions or inclusion criteria about the properties of charts in a corpus. These are usually specified to constrain the research problem space to achieve feasible solutions~(e.g., ChartReader~\cite{rane2021chartreader} requires the input to be bar charts). 
Based on our paper coding, the scope of a corpus is primarily defined through \textit{chart types}. In addition, we have also identified additional assumptions on the extent of \textit{design variations within a chart type}. 
In the following subsections, we summarize researchers' common choices and practices along these two dimensions.

\subsection{Chart Type}
High-level chart typologies are commonly used to define the scope of a research problem (hence the scope of a corpus). For example, Gao~\etal~\cite{gao2012view} and Choi \etal~\cite{choi2019visualizing} extract chart semantics such as label positions, chart type and source data from three types of charts: bar, line, and pie; DVQA~\cite{kafle2018dvqa} and Chartreuse \cite{cui2021mixed} focus on reusing bar charts. 
Table~\ref{tbl:typeFrenquency} presents the frequency of each chart type used in our collected papers. 
Our analysis does not include \cite{chen2020composition,wu2020mobilevisfixer,hoque2019searching,oppermann2020vizcommender} because the first two only included visual layout and scale requirements, the third one works at a level of granularity finer than high-level chart typologies, and the last one didn't reveal information regarding chart types.



\begin{table}[h]
    \centering
    \caption{Used frequency and percentage~(calculated by dividing the frequency by 52 since 4 corpora were not counted) of each chart type in surveyed chart corpora.
    }
    \begin{tabular}{lcr}
    \toprule
     Chart Type    &  Frequency &  Percentage\\
     \midrule
      Bar                 & 38 & 73.07\%\\
Line                & 31 & 59.62\% \\
Pie                 & 18 & 34.62\% \\
Scatterplot         & 16 & 30.77\% \\
Infographics        & 9  & 17.31\% \\
Area                & 9  & 17.31\% \\
Map                 & 8  & 15.38\% \\
Treemap             & 4  &  7.69\% \\
Boxplot             & 4  & 7.69\% \\
Heatmap             & 3  & 5.77\% \\
Table               & 3  & 5.77\% \\
Venn                & 3  & 5.77\% \\
Parallel Coordinate & 3  & 5.77\% \\
Sunburst            & 3  & 5.77\% \\
Donut               & 3  & 5.77\% \\
Node-link Diagram   & 3  & 5.77\% \\
Radar               & 2  & 3.85\% \\
Matrix              & 2  & 3.85\%\\
Tick                & 2  & 3.85\%\\
Pareto              & 2  & 3.85\%\\
     \bottomrule
    \end{tabular}
    \label{tbl:typeFrenquency} 

\end{table}

From Table~\ref{tbl:typeFrenquency}, we can see that among the many chart types observed, bar, line, pie charts, and scatterplots are the most frequent ones, which is consistent with the statistics from the Beagle dataset~\cite{battle2018beagle} where line and bar charts dominate its collections and pie chart is less popular. The popularity of these chart types can be attributed to their effectiveness in visualizing numerical data~\cite{kafle2018dvqa}, the ability to convey trends and relationships~\cite{rane2021chartreader} and to represent high-dimensional data in 2D~\cite{ma2018scatternet}, as well as their simplicity and interpretability~\cite{battle2018beagle}. Infographics, maps, and area charts, typically of specific usages, are less frequently included and appear in about $17\%$ of all the corpora, respectively. The other chart types, such as donut charts and tick plots, are rarely considered, potentially due to their relatively scarce presence on the web and increased visual structure complexity.


\begin{table*}[ht]
    \centering
    \caption{A categorization of scope regarding design variations observed in collected corpora. The three columns are high-level design variation types, low-level details assumptions over visual designs, and corresponding chart corpora, respectively.}
    \begin{tabular}{lll}
    \toprule
     Design Variation Type   &  Assumption &  Relevant Corpora\\
     \midrule
\multirow{3}{*}{composite arrangement}                & only  multiple-view charts & \cite{chen2020composition} \\
             & no  multiple-view charts & \cite{chagas2018evaluation,wu2020mobilevisfixer,poco2017reverse,hsu2021scicap,lu2020exploring} \\
             & no layered charts & \cite{jung2017chartsense, poco2017reverse,choi2019visualizing} \\
             \midrule
\multirow{4}{*}{mark and glyph}    & no abstract icons or symbols & \cite{jung2017chartsense} \\
               & only proportion-related charts  & \cite{qian2020retrieve,cui2019text} \\
               & only timeline-related infographics  & \cite{chen2019towards} \\
           & no handmade sketches & \cite{jung2017chartsense,chagas2018evaluation,saleh2015learning} \\
             & no 3D effects & \cite{savva2011revision,deng2022visimages,choi2019visualizing} \\
             \midrule
\multirow{2}{*}{chart component}                 & chart must have a legend & \cite{poco2017extracting,mahinpei2022linecap} \\
                 & axes being at the left and bottom & \cite{savva2011revision} \\
                 \midrule
  coordinate space               & in Cartesian coordinate space & \cite{wu2020mobilevisfixer, poco2017reverse} \\
     \bottomrule
    \end{tabular}
    \label{tbl:designConstraint}
\end{table*}

\begin{figure}[htbp]
\centering
\includegraphics[width=0.475\textwidth]{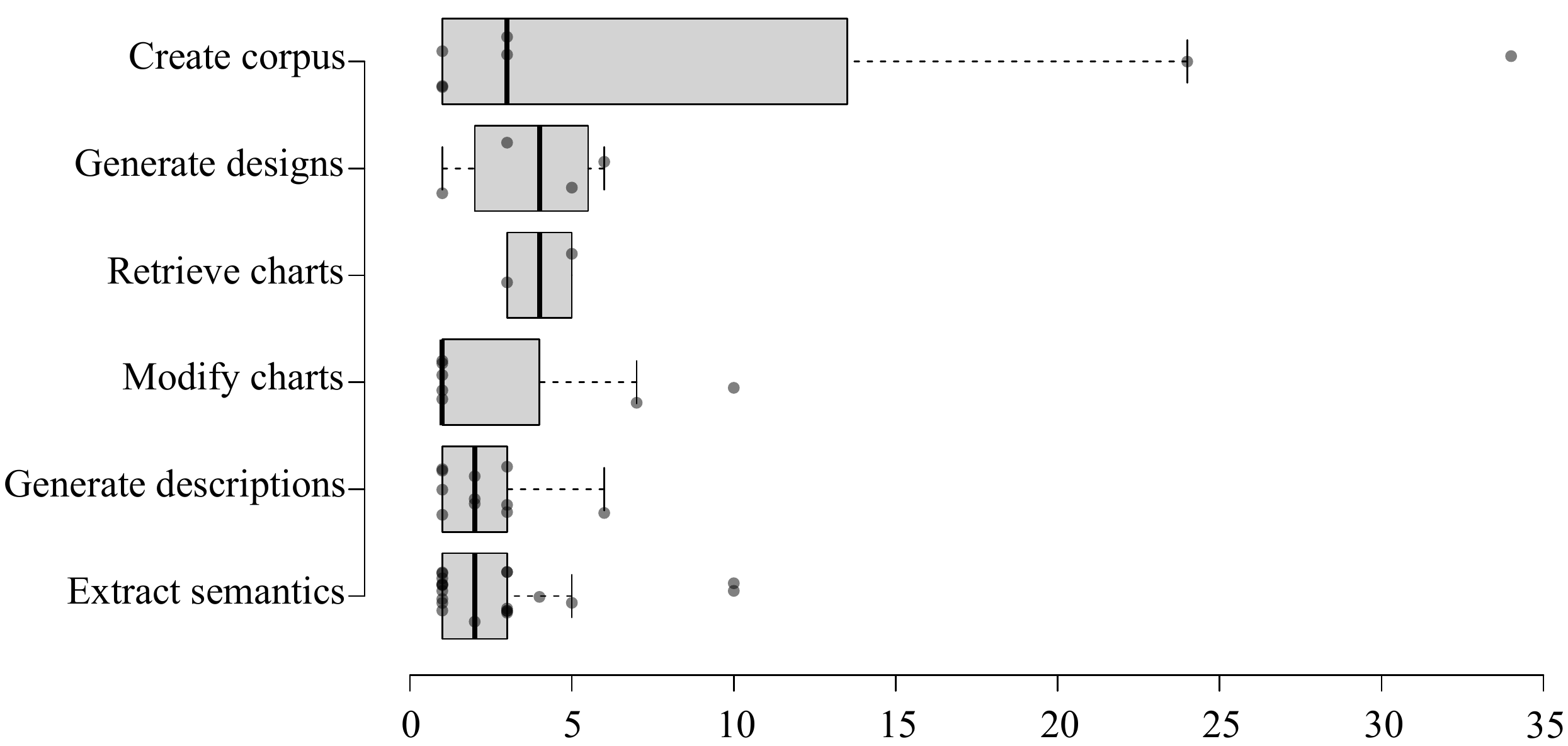}
\caption{The distribution of the number of chart types in collected corpora by task goal.}
\label{fig:typeVsPurposBoxplot}
\end{figure}

The intrinsic nature and complexity of the task can also influence the number of chart types in a corpus. 
Figure~\ref{fig:typeVsPurposBoxplot} shows the distribution of the number of chart types by task purpose in jittered box plots, from which we can see most corpora contains fewer than 10 chart types.
Unsurprisingly, \textit{create a chart corpus} leads to the highest average number of chart types, since this task typically aims at a comprehensive and diverse chart collection. Two runner-ups following~\textit{create a chart corpus} are \textit{generate chart designs automatically}, which oftentimes employs end-to-end learning-based models that don't require type-specific handling~(e.g., all four works, Zhao~\etal~\cite{zhao2020chartseer}, Dibia and Demiralp~\cite{ dibia2019data2vis}, Cui~\etal~\cite{cui2019text}, and Hu~\etal~\cite{hu2019vizml}, that belong to this task used neural networks as the pivot method), and \textit{retrieve charts matching certain criteria}, which may not require a deep understanding of chart semantics~(DiagramFlyer~\cite{chen2015diagramflyer} is a search engine for charts primarily based on label text in axes and legends). It is thus possible to handle more chart types in these works. 
Each of the other three purposes --- \textit{extract chart semantics}, \textit{modify an existing chart
}, and \textit{generate natural language descriptions
} --- has a relatively small average number of chart types. These tasks usually require a more thorough chart deconstruction and understanding of the chart semantics; thus, involving too many chart types might make the research problem intractable~\cite{jung2017chartsense,luo2021hybrid}. For example, Chartreuse~\cite{cui2021mixed} developed algorithmic heuristics to obtain element groupings, data-binding regimes, spatial layouts, and approximated underlying data from infographics bar charts; MapQA ~\cite{chang2022mapqa} first adopted OCR techniques to extract text elements in a chart, then used neural networks to recover underlying data and synthesize possible answers to questions solely for choropleth maps.

In practice, we have also observed that different corpora were created for the same chart analysis task, due to different scopes. 
For example, Cui~\etal~\cite{cui2021mixed} collected their own infographics bar chart corpus instead of reusing the corpus from Chen~\etal~\cite{chen2019towards}, which focused on timeline infographics. Although their tasks are both \textit{modify an existing chart}, their scopes diverge, requiring different corpora. This leads to challenges on the generalizability/scalability of presented techniques/systems. 

\begin{figure}[htbp]
\centering
\includegraphics[width=0.475\textwidth]{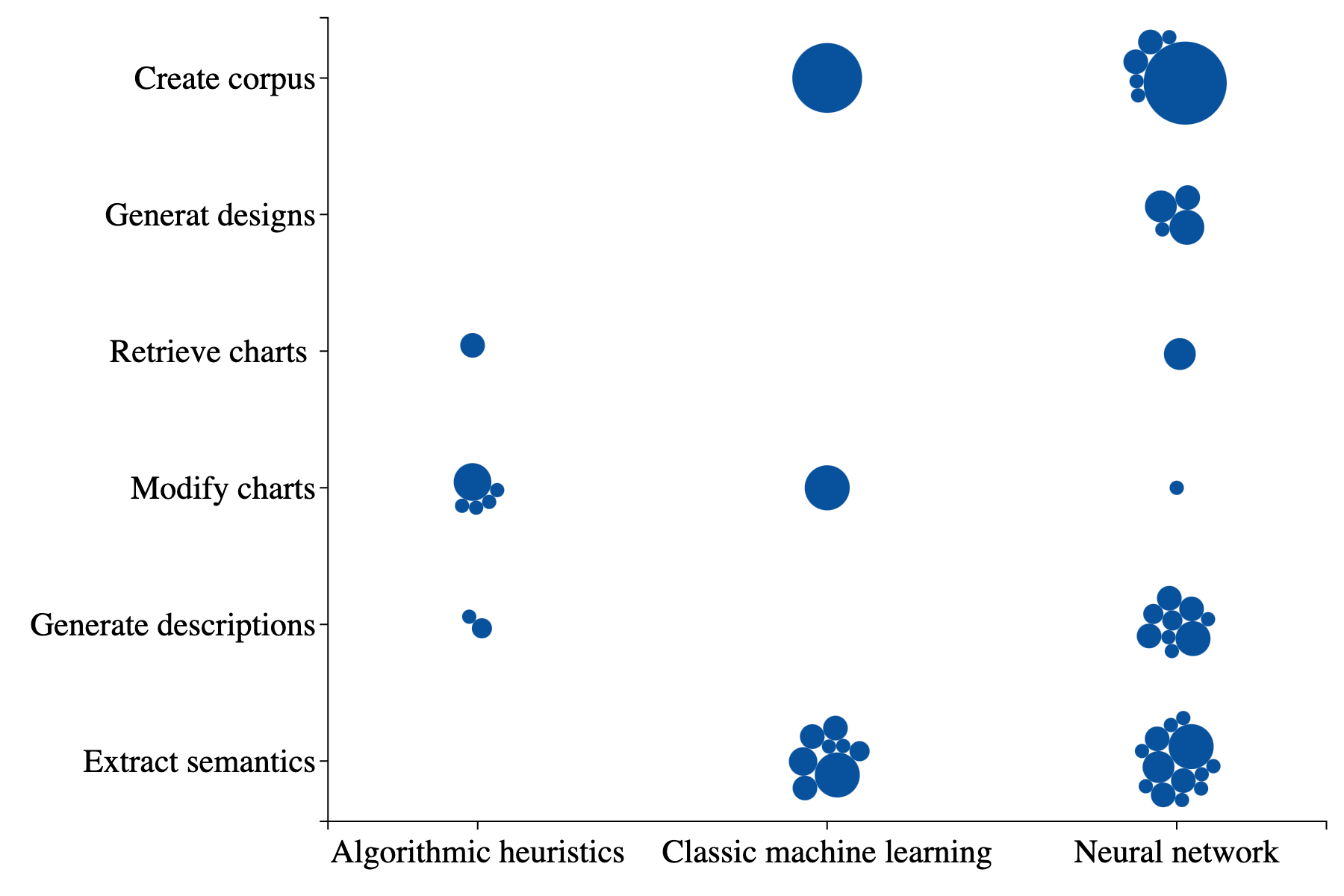}
\caption{The clusters of chart corpora based on task goal and method. Each dot represents a chart corpus whose size encodes the number of chart types considered.}
\label{fig:typeVsPurposeMethod}
\end{figure}

We also examine how the methods play a role in influencing the number of chart types in a corpus. 
Figure~\ref{fig:typeVsPurposeMethod} shows a beeswarm chart presenting the clusters of chart corpora based on task goal and method, where each dot represents a corpus whose size encodes the number of chart types considered. It can be seen that the corpora used for neural networks or classic machine learning models tend to contain more chart types, which is possibly due to the higher representational capacities of the two methods compared to heuristics-based algorithms. 

\vspace{-0.12in}

\subsection{Design Variations}
In addition to chart type, we have also observed scope definitions in terms of finer-grained design variations in some corpora. There can be different structural and stylistic variations within a chart type, and supporting all these variations is non-trivial. Examples of design variations include but are not limited to: 
composite arrangement (e.g., Chen \etal~\cite{chen2020composition} focus on decomposing and understanding multiple-view visualizations, and Poco and Heer~\cite{poco2017reverse} assume non-superimposed single-layer figures), and mark/glyph type (e.g., Chen \etal~\cite{chen2019towards} retarget timeline glyphs in infographics).
We summarize scope definitions related to design variations in Table~\ref{tbl:designConstraint}. Although the assumptions on design variations enforced on a corpus may not be explicitly described in some papers, these constraints are essential during the corpus curation process to filter out undesired charts and keep the research focus manageable ~\cite{lai2020automatic}.

%% file: sections/6_collection.tex
\section{Chart Collection Method}

The collection method describes how the charts in a corpus were collected. The choice of method is determined by both the chart format and the corpus scope. During our coding process, we have observed four kinds of collection methods: \textit{reusing and transforming existing corpus}~\cite{zhao2020chartseer,chen2019figure,bylinskii2017learning,singh2020stl}, \textit{web crawling}~\cite{chen2015diagramflyer,battle2018beagle,hoque2019searching,hu2019vizml,li2022structure}, \textit{manual curation}~\cite{huang2007system, gao2012view,demir2012summarizing, poco2017extracting,cui2019text,choi2019visualizing,qian2020retrieve}, and~\textit{computer-aided generation}~\cite{sharma2019chartnet,chaudhry2020leaf,methani2020plotqa,cliche2017scatteract}.
In the following subsections, we describe these collection methods and summarize the common sources and tools people adopted during the curation process.
Note that these four methods are not mutually exclusive: one can combine multiple methods to create a corpus. For example, ChartSense~\cite{jung2017chartsense} reused the Revision corpus~\cite{savva2011revision} and augmented it with more web-crawled images.


\subsection{Reusing and Transforming Existing Corpus}
\label{sec:reuse_corpus}
Directly reusing existing chart corpora  is straightforward and requires minimal effort. 
However, out of the 56 corpora described in our paper collection, 17 are publicly available~(which is consistent with the observation from Davila~\etal~\cite{davila2020chart} that ``very few of the datasets have been made publicly available''), 9 were generated by modifying existing corpora, and only 4 corpora~(FigureQA~\cite{kahou2017figureqa}, VIF~\cite{lu2020exploring}, SciCap~\cite{hsu2021scicap}, REV~\cite{poco2017reverse}) were reused in subsequent works.
This shows a lack of standard benchmark corpora in visualization research, as discussed in the introduction. 
Two kinds of transformations are applied to those 9 corpora that were built by modifying existing corpora:
\begin{itemize}
    \item Adding new charts to an existing corpus to make a larger one. For example, the corpora in~\cite{cui2021mixed,zhao2020chartseer,jung2017chartsense,kim2018multimodal} are created by augmenting corpora from~\cite{madan2018synthetically, dibia2019data2vis,savva2011revision,wu2012recognizing}, respectively. The motivation behind this augmentation is either simply increasing the corpus size~\cite{jung2017chartsense} or increasing the chart diversity~\cite{zhao2020chartseer}. The methods they used to obtain new charts include \textit{manual collection}, \textit{web crawling}, and~\textit{computer-aided generation}, which will be introduced in later subsections. 
    \item Adding new annotations to the same charts. For example, the corpora in~\cite{chen2019figure,singh2020stl,hsu2021scicap} are created by adding new question-caption~(QC) or question-answer~(QA) annotations on the corpora from~\cite{kahou2017figureqa,chaudhry2020leaf,clement2019use}, respectively; Bylinksii~\etal~\cite{bylinskii2017learning} and Fu~\etal~\cite{fu2019visualization} built their corpora by annotating salience map and aesthetics score respectively on the MassVis dataset~\cite{borkin2013makes}.
\end{itemize}

\subsection{Web Crawling}
To quickly collect a large chart corpus, 
web crawling is a popular way to gather 
charts matching certain criteria from targeted sources automatically. We have observed the following commonly used websites
that people add to their crawlers:

\begin{enumerate}
    \item Search engines in which people conduct keyword-based searches; e.g., Google Search~\cite{GoogleImage} used in~\cite{jung2017chartsense,wu2020mobilevisfixer, savva2011revision, choi2019visualizing}.
    
    \item Galleries of online charting tools, e.g., Tableau~\cite{Tableau} used in~\cite{oppermann2020vizcommender}, Plotly~\cite{Plotly} used in~\cite{battle2018beagle,hu2019vizml,li2022structure}, Chartblocks~\cite{Chartblocks}, Fusion Charts~\cite{Fusioncharts}, and Graphiq~\cite{Graphiq} used in~\cite{battle2018beagle}, and D3~\cite{bostock2011d3} used in~\cite{battle2018beagle,wu2020mobilevisfixer,hoque2019searching}.
    
    \item Public documented materials, e.g., online Excel sheets used in~\cite{luo2021chartocr, luo2021hybrid, huang2021visual}.

    \item Public scholarly document repositories, examples are Vispubdata~\cite{vispubdata}, DBLP~\cite{Dblp}, Semantic Scholar~\cite{SemanticScholar}, ACL Anthology repository~\cite{Acl}, and CiteSeerX repository~\cite{CiteSeerX} used in~\cite{chen2020composition,chen2020composition,poco2017extracting,poco2017reverse,choudhury2016scalable}, respectively.

    \item Public platforms for sharing data analysis and reports, examples are Statista~\cite{Statista} used in~\cite{obeid2020chart,masry2022chartqa}, the Pew research~\cite{Pew}, Our World In Data~(OWID)~\cite{Owid}, and Organisation for Economic Co-operation and Development~(OECD)~\cite{Oecd} used in~\cite{masry2022chartqa}, Kaiser Family Foundation~(KFF)~\cite{Kff} used in~\cite{chang2022mapqa}, and Quartz~\cite{Quartz} used in~\cite{poco2017reverse}.
\end{enumerate}

Although web crawlers can gather a large number of charts, 
it is also mentioned that web crawling usually requires some manual post-examination to remove the repeated or unqualified charts~\cite{chen2019towards,poco2017reverse,chen2020composition,jung2017chartsense} from a typically large chart collection, which requires additional time. The variety of charts in terms of types and design variations cannot be guaranteed either.

\subsection{Manual Curation}\label{sec:manualCollecting}
Due to the above-mentioned limitations of web crawling, when the quality and variation of chart design matters more than the size of a corpus, some works, e.g., Cui~\etal~\cite{cui2019text} and Chen~\etal~\cite{chen2019towards}, 
decided to collect charts manually without the help of (semi-)automated crawlers, so that they could inspect each chart candidate and decide if they would like to include it in the corpus. We identify three common kinds of sources where people find charts manually:
\begin{enumerate}
    \item Search engines in which people adopt keyword-based searches; candidates include Google Search~\cite{GoogleImage} used in~\cite{lai2020automatic,chagas2018evaluation,qian2020retrieve,zhou2021reverse}, Bing Visual Search~\cite{Bing} and Yahoo Image Search~\cite{Yahoo} used in~\cite{zhou2021reverse}, Freepik~\cite{Freepik} and ShutterStock~\cite{Shutterstock} used in~\cite{lu2020exploring}, Flickr~\cite{Flickr} used in~\cite{saleh2015learning}~(the latter three contain more infographics).
    \item Galleries of online charting tools, e.g., Vega-Lite gallery~\cite{satyanarayan2016vega,Vegalitegallery} and D3 gallery~\cite{bostock2011d3,D3Gallery} used in~\cite{kim2020answering}.
    \item Published or publicly available materials, examples are academic papers~\cite{al2017machine,hsu2021scicap,kim2020answering,rane2021chartreader,deng2022visimages}, online PowerPoint templates~\cite{cui2021mixed,cui2019text}, and paper media such as magazines and newspapers~\cite{demir2012summarizing}.
\end{enumerate}

In both manual curation and web crawling, when the chart source are scholarly document repositories, necessary post-processing is needed to further extract charts from collected academic papers in the PDF format. The commonly used tools include PDFFigures~\cite{clark2015looking,clark2016pdffigures} used in~\cite{al2017machine,rane2021chartreader,poco2017extracting,poco2017reverse,choudhury2016scalable,hsu2021scicap,deng2022visimages}, PyMuPDF~\cite{Pymupdf} and PDF2HTML~\cite{Pdf2html} used in~\cite{chen2020composition}, and Diagram Extractor~\cite{chen2011searching} used in~\cite{chen2015diagramflyer}.

Note that one can combine multiple sources to perform manual curation to increase corpus size and enhance diversity~(we discuss chart diversity in detail in Section~\ref{sec:diversity}). 

\subsection{Computer-Aided Generation}\label{computerAidedGeneration}

Another method for preparing a chart corpus is computer-aided generation, i.e., using visualization software to generate charts based on real or synthetic datasets. Three questions need to be addressed when using this method: Where do the underlying datasets come from? Which charting tools to use? How to ensure a wide range of design variations and styles?

\bpstart{Underlying datasets.} Two ways of preparing underlying datasets for plotting are observed: (1) using synthetic datasets generated by computers with various data types~\cite{sharma2019chartnet}, random-distributed values~\cite{kafle2018dvqa,chagas2018evaluation,cliche2017scatteract} or values from carefully-tuned distributions~\cite{kahou2017figureqa,chang2022mapqa}, and (2) using public data tables available from various online sources~\cite{chaudhry2020leaf,methani2020plotqa,ma2018scatternet, chen2019towards}, e.g., World Development Indicators~\cite{Wdi}, Historical daily prices and volumes of all U.S. stocks and ETFs~\cite{Stockdata}, and the PyDataset library~\cite{Pydataset}.

\bpstart{Charting tools.} The most popular charting tool in our collection is Matplotlib~\cite{Matplotlib} from python, which is used in 5 corpora~\cite{kafle2018dvqa,chaudhry2020leaf,zhou2021reverse,cliche2017scatteract,ma2018scatternet}. Other tools include Vega-Lite~\cite{satyanarayan2016vega,Vegalitegallery} used in~\cite{chagas2018evaluation}, Bokeh~\cite{Bokeh} used in~\cite{kahou2017figureqa}, GeoPandas~\cite{Geopandas} used in~\cite{chang2022mapqa} for map-based charts, and TimelineStoryteller~\cite{Timelinestoryteller} used in~\cite{chen2019towards} for timeline-based infographics.

\bpstart{Enhancing design variety.} Unlike manual curation and web crawling, to generate charts using computers, one has to enforce design variety when using charting tools. Two common practices have been observed: increasing the diversity of (1) underlying datasets and (2) visual styles. We detail this discussion in section~\ref{sec:diversity}.


\begin{figure}[htbp]
\centering
\includegraphics[width=0.475\textwidth]{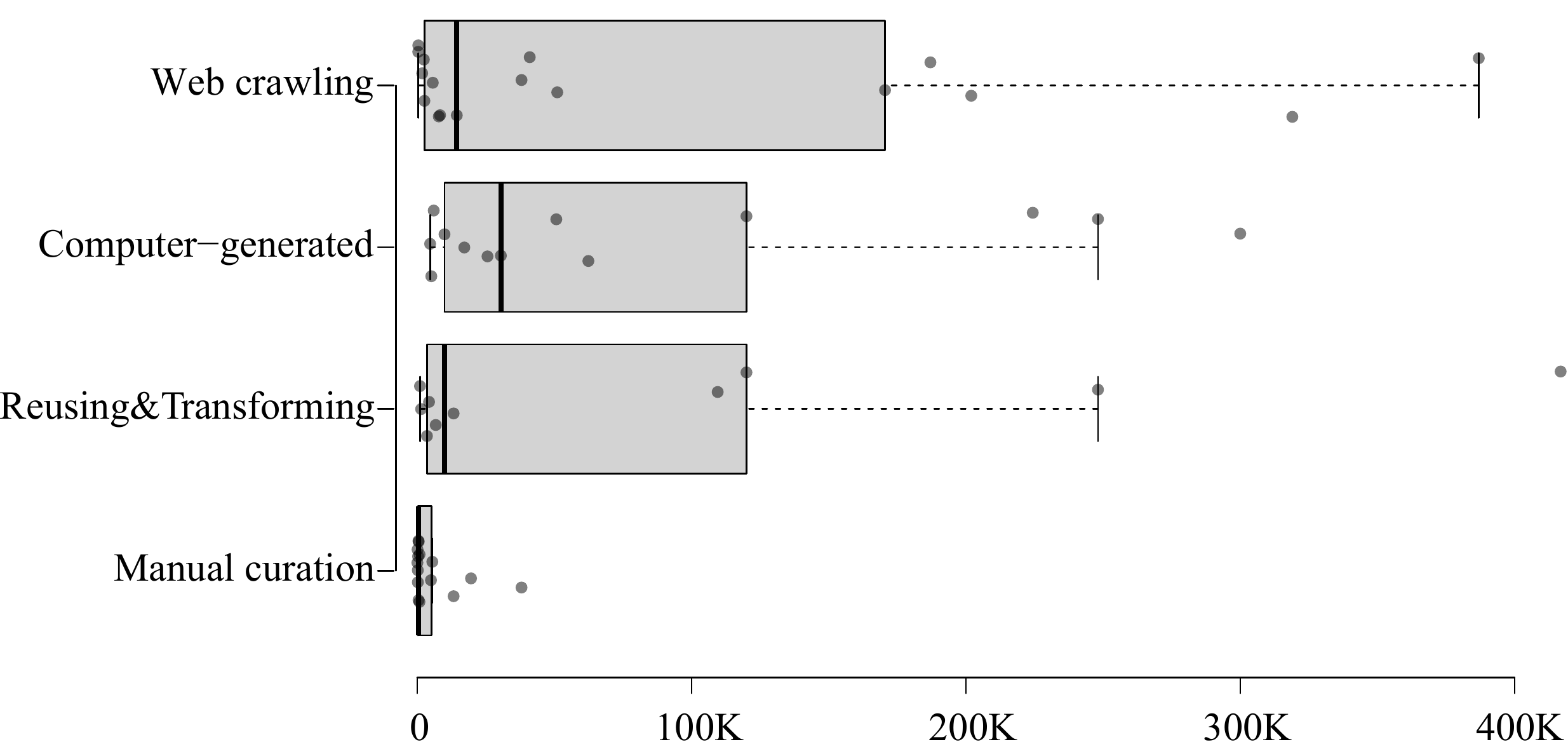}
\caption{The distribution of corpus size by chart collection method.}
\label{fig:sizeVsCollectionBoxplot}
\end{figure}

Figure~\ref{fig:sizeVsCollectionBoxplot} shows the distribution of corpus size across different collection methods. The corpus from~\cite{hu2019vizml} is not included since its corpus size is too big to make the figure readable; in cases where multiple collection methods were used, we choose the one accounting for the largest portion. It can be seen that, on average, web crawling and computer-aided generation lead to corpora of large sizes, and manual curation unsurprisingly results in small-size corpora. 

%% file: sections/7_annotation.tex
\section{Annotations}

Annotations are labels associated with charts in a corpus, serving as ground truth for chart analysis tasks. In most cases, the sources where the charts are collected do not provide such labels. Also, as reported in Battle~\etal~\cite{battle2018beagle}, there is a lack of consistent metadata across different websites, which makes automatic labeling hard. For example, Plotly~\cite{Plotly} and Chartblocks~\cite{Chartblocks} provide chart type information while Fusion Charts~\cite{Fusioncharts}, Graphiq~\cite{Graphiq}, and D3~\cite{bostock2011d3} do not. Thus, it is necessary to annotate collected charts to obtain consistent valid labels for a given task. In cases where the sources contain meta-information about the charts, the provided information is not always sufficient for the task. For example, the chart-type information provided by Plotly~\cite{Plotly} is not enough for tasks like \textit{modify an existing chart} since this task usually requires knowing more low-level semantics. 

In this section, we discuss two aspects of annotations: \textit{annotation types}, which refer to the categories of labels needed for a variety of tasks,
and \textit{annotation methods}, which refer to the approaches people adopt to obtain the desired labels.

\subsection{Annotation Types}\label{sec:annotationType}

\begin{table*}[!htbp]
    \centering
    \caption{Typical annotation types in the collected chart corpora.}
    \begin{tabular}{lp{0.7\linewidth}}
    \toprule
     Annotation Type    &  Relevant Corpora\\
     \midrule
bounding box                &  for mark or glyph~\cite{lai2020automatic,luo2021chartocr,luo2021hybrid,cliche2017scatteract,chaudhry2020leaf,qian2020retrieve,huang2021visual}, \\ & for legend~\cite{luo2021hybrid,chaudhry2020leaf,methani2020plotqa,huang2021visual}, for axes~\cite{methani2020plotqa,cliche2017scatteract,huang2021visual}, for text~\cite{luo2021hybrid,zhou2021reverse,poco2017reverse,cliche2017scatteract,sharma2019chartnet,chaudhry2020leaf,methani2020plotqa,savva2011revision,huang2021visual}, for main chart area~\cite{luo2021chartocr,deng2022visimages,huang2021visual}, for chart sub-views~\cite{chen2020composition} \\
             \midrule
chart type    & \cite{battle2018beagle,jung2017chartsense,tang2016deepchart,chagas2018evaluation,kim2018multimodal,savva2011revision,choudhury2016scalable,gao2012view,deng2022visimages,choi2019visualizing} \\ \midrule
question-answer pair & \cite{kim2020answering,masry2022chartqa,kafle2018dvqa,kahou2017figureqa,mathew2022infographicvqa,chaudhry2020leaf,chang2022mapqa,methani2020plotqa} \\  \midrule
question-caption pair & \cite{chen2019figure,mahinpei2022linecap} \\  \midrule
text role    & \cite{zhou2021reverse,choudhury2016scalable,poco2017reverse} \\ \midrule
infographics element type    & \cite{lu2020exploring,qian2020retrieve} \\ \midrule
pairwise style similarity & \cite{saleh2015learning,ma2018scatternet} \\  \midrule
 saliency map  & \cite{bylinskii2017learning} \\  \midrule
  aesthetics ranking  & \cite{fu2019visualization} \\
     \bottomrule
    \end{tabular}
\label{tbl:annotationType}
\end{table*}

In Table~\ref{tbl:annotationType}, we summarize typical annotation types observed in our analysis. It can be seen that bounding box annotation for chart elements is the most common one since knowing the positions of certain chart elements is necessary across most tasks. For example, Deng~\etal~\cite{deng2022visimages} annotated the bounding boxes of the main chart areas to record accurate visualization locations which serve as one of the output components; Huang~\etal~\cite{huang2021visual} annotated the bounding boxes of legends to develop an object detection model that predicts locations of legends in new charts; Poco and Heer~\cite{poco2017reverse} annotated bounding boxes of text elements to test their OCR-based technique for locating and extracting text content; Chen~\etal~\cite{chen2020composition} annotated bounding boxes of sub-views to advance their chart composition and configuration analysis as well as to develop the recommendation system that retrieves charts with a similar layout; and Chaudhry~\etal~\cite{chaudhry2020leaf} annotated bounding boxes for a variety of chart elements like axis labels, legend, and marks to train their Mask-RCNN network~\cite{he2017mask} for chart element detection and classification, which will later be utilized to develop question answering models. We can see that bounding box annotations are the foundation of many different tasks and models, even when element positions are not required in the final output.

Many corpora also include chart type annotations, with which people can perform chart type classification tasks~\cite{battle2018beagle,deng2022visimages}.
Chart type annotations further allow deeper chart deconstruction and understanding. For example, Jung~\etal~\cite{jung2017chartsense} first trained predictive models to classify charts, then extracted underlying source data per chart type; chart type classification is also a prerequisite in the Revision system~\cite{savva2011revision} for redesigning an existing chart.

Question-Answer~(QA) pair annotation is also commonly seen due to the increasing popularity of chart question-answering systems~\cite{sharma2019chartnet,chaudhry2020leaf,demir2012summarizing,hsu2021scicap,chen2019figure}.
Taking a single bar chart as an example, questions that can be asked generally have the following types: (1) structure-related~\cite{kafle2018dvqa}, such as \textit{are the bars horizontal?}, (2) data-related~\cite{kim2020answering}, such as \textit{what is the label of the third bar?}, and (3) relation-related~\cite{chang2022mapqa}, such as \textit{what are the highest and lowest values?}.
A similar annotation type is Question-Caption~(QC) pair, observed in chart captioning systems~\cite{chen2019figure,mahinpei2022linecap}. Both of them are only considered in the task of \textit{generate natural language descriptions}, where answers to questions or captions to charts are outputs of the type~\textit{synthesized descriptions} shown in Section~\ref{sec:task}.

Some rarely-seen annotation types are observed in specific corpora for task needs:
saliency map~\cite{bylinskii2017learning}, aesthetics ranking~\cite{fu2019visualization}, text orientation~\cite{savva2011revision}, x-axis labels~\cite{obeid2020chart}, infographics-specific attributes such as timeline-based representations, scales and layouts~\cite{chen2019towards}, and color-text correspondence~\cite{poco2017extracting}.

\subsection{Annotation Methods}\label{sec:annotationMethod}
We identify four kinds of methods for obtaining annotations:
\begin{itemize}
    \item \textbf{In-house labeling}: an in-person annotation process where a small group of people gathers together to annotate collected charts manually. This method is commonly used and usually works for datasets of relatively small sizes. Two issues need to be considered when performing in-house labeling:\begin{enumerate}
        \item User interface for annotation. 
        The choice of user interface 
        depends on the complexity of the annotation type.
        For example, for relatively simple annotations like chart types~\cite{battle2018beagle,choudhury2016scalable}, a graphical user interface may not be necessary. For annotations that require high accuracy or repeated manual operations~(e.g., specifying bounding boxes for sub-views~\cite{chen2020composition}, and the position, size, angular orientation, and content of text regions~\cite{savva2011revision}), a graphical user interface can facilitate the labeling process and improve the quality of annotations. 
        \item Training procedure. To help annotators better understand the annotation types and tasks, select qualified annotators, and increase the annotation quality, training is typically required. For example, Deng~\etal~\cite{deng2022visimages} carried out a training session before the formal annotation process, which covered the details of annotation types and tasks; they later asked participants to finish a test based on the introduction and used the test results to identify eligible annotators. 
    \end{enumerate}
    In-house labeling can be impractical when the corpus size is large and the required annotation types are complex~\cite{lu2020exploring}; alternative approaches shall be considered in those cases. 
    \item \textbf{Crowdsourcing}: an online annotation process where workers from crowdsourcing platforms such as Amazon’s Mechanical Turk are recruited to annotate charts. This method usually is considered when the size of a chart corpus is large. 
    
    The two considerations described for in-house labeling also apply here: Crowdsourcing often involves a GUI for labeling and a training session to teach online workers the annotation tasks~(e.g., Saleh \etal~\cite{saleh2015learning} included three examples in the introduction session to teach the workers the purpose of the experiment as well as the meaning of the annotation type - stylistic infographics similarity) and identify qualified workers~\cite{mahinpei2022linecap}. Besides, a formal post-examination is usually carried out by the organizers to remove invalid annotations, because the quality of annotations from online workers varies even if a training session is included. For example, Kim \etal~\cite{kim2020answering} manually reviewed the annotations and removed QA pairs that were not reasonable given the charts.

    It is worth mentioning that despite the possibility of annotating large-scale corpora through crowdsourcing, the cost of hiring online workers can be high in practice~\cite{methani2020plotqa,saleh2015learning}, which makes it not always the first choice or a feasible option for annotating large corpora.

    \item \textbf{Template-based generation}: annotations in the form of QA or QC pairs for given charts can be generated based on pre-defined templates. This approach is observed solely in corpora built for the purpose to~\textit{reason about communicative information}~\cite{sharma2019chartnet,kafle2018dvqa,chen2019figure,kahou2017figureqa,chaudhry2020leaf,chang2022mapqa}. Compared to crowdsourcing, template-based QA/QC generation avoids high expenses, but in general lack rich linguistic variations~\cite{masry2022chartqa}. Some works alleviate this diversity issue by developing more sophisticated templates~\cite{chen2019figure,singh2020stl}, combining template-based generating with crowdsourcing~\cite{methani2020plotqa}, or adopting large-scale language models~\cite{masry2022chartqa}. Section~\ref{sec:diversity} discusses more details on how to diversify such annotations.

    \item \textbf{Automatic extraction} is applicable when the corpus is generated computationally  or collected from Excel sheets. 
    For example, the bounding boxes of chart elements in the corpora can be extracted using Matplotlib if the charts were generated using the same tool~\cite{zhou2021reverse,cliche2017scatteract,methani2020plotqa}; the bounding boxes and underlying data values can be extracted using meta-information of charts generated using Excel~\cite{luo2021chartocr,luo2021hybrid,huang2021visual}.
    When the tools and associated code that generated a corpus are not available, 
    automatically extracting labels requires building chart-to-label models, but the extraction results may not be satisfying: e.g., Chen \etal~\cite{chen2020composition} developed a YOLOv3 object detection model~\cite{redmon2018yolov3} to segment a given chart into sub-views automatically. However, the model's output performance metric was not accurate enough, forcing the authors to label the sub-views manually.
    
\end{itemize}

In addition to the above four annotation methods, hiring a professional data annotation company was used in the creation of VisImage corpus~\cite{deng2022visimages} for bounding box annotations. This is an expensive method~\cite{saleh2015learning}, thus is rarely considered.











%% file: sections/8_diversity.tex
\section{Chart Diversity}\label{sec:diversity}

Diversity measures how much the visualizations differ from one another within a corpus. This integrative property depends on factors including chart format, scope, and collecting method. For example, Deng~\etal~\cite{deng2022visimages} demonstrated better diversity by showing a more balanced distribution over chart types compared to the MassVis dataset~\cite{borkin2013makes}; Li \etal~\cite{li2022structure} acknowledged source diversity 
as one limitation 
since their system was trained with SVG images crawled solely from Plotly~\cite{Plotly} and may 
not work well on 
visualizations created with other tools or from other sources. In general, diversity is an important property that could significantly influence the scalability, generalizability, and robustness of developed techniques or systems, and it is under-explored in the current literature compared to other properties. In this section, we summarize current practices to enhance diversity in chart corpora.



\bpstart{Diversify source websites.} The most common and straightforward way to enhance diversity is to collect charts from multiple sources, varying source websites typically lead to greater chart design variations. This approach works for both manual curation and web crawling. For example, Wu~\etal~\cite{wu2020mobilevisfixer} built a web crawler based on a D3 chart crawler~\cite{hoque2019searching}, augmented the seeding pages with sources like Google Image Search~\cite{GoogleImage}, and randomized the visiting queue of their crawler to avoid human bias and further increase diversity. In InfographicVQA~\cite{mathew2022infographicvqa}, the corpus contains infographics downloaded from thousands of different websites, with a variety of visual designs. It is thus more diverse than previous infographics corpora which were either specialized collections~(e.g., the MassVis corpus~\cite{borkin2013makes} focuses on the illustration of scientific procedures and statistical charts~\cite{mathew2022infographicvqa}) or obtained from one single source.

\bpstart{Diversify chart topics.} Besides adding more sources websites, collecting charts on various topics is another way to promote diversity since datasets and suitable visual designs vary by topic. In practice, there are mainly two ways to diversify chart topics:
\begin{enumerate}
    \item Including diverse topics in search keywords.
    For example, the chart corpus in Retrieve-Then-Adapt~\cite{qian2020retrieve} was created by combining a primary keyword, “infographic”, with numerous secondary keywords indicating 10 different topics~(\eg ``education'', ``health'', and ``commerce'') to ensure diversity, and ended up containing 1000 infographic sheets with 100 under each topic. 
    \item Adopting topic-enriched source websites, which means sampling charts from websites containing  rich and diverse content topics. For example, Masry~\etal~\cite{masry2022chartqa} used Statista~\cite{Statista}, a public platform that presents charts covering various topics such as economy, politics, and industry, as one of their collecting sources; the MassVis corpus reused in Bylinskii~\etal~\cite{bylinskii2017learning} included charts from several websites including WHO~\cite{Who} and Wall Street Journal~\cite{Wallstreet} to span a diverse set of topics including public health, economy, and public policy.
\end{enumerate}

\bpstart{Diversify chart creators.} Some corpora sample from online image providers where charts are created by a larger community of content creators. This strategy leverages the variety of real-world users' datasets and chart design preferences, thus promoting diversity. For example, Lu \etal~\cite{lu2020exploring} collected charts from two providers, Shutterstock~\cite{Shutterstock} and Freepik~\cite{Freepik}, because infographics from these sites are contributed by designers worldwide, and span a variety of design themes and styles; Li~\etal~\cite{li2022structure} and Hu~\etal~\cite{hu2019vizml} sampled user-created charts from Plotly~\cite{Plotly} and kept one chart per user.

\bpstart{Diversify scholarly document repositories.} When targeting charts from scholarly documents for scientific purposes, enriching publication venues and increasing the year range are standard practices to increase diversity. 
For example, the VisImage corpus~\cite{deng2022visimages} collected charts from 22-year VAST and InfoVis publications; the MV dataset~\cite{chen2020composition} is created using publications in IEEE VIS, EuroVis, and IEEE PacificVis from 2011--2019; Choudhury~\etal~\cite{choudhury2016scalable} chose papers in top 50 computer science conferences from 2004 to 2014. Some other strategies people use to sample scientific charts include: increasing publication fields~(\eg Poco \etal~\cite{poco2017extracting} selected charts from five areas including visualization, human-computer interaction, computer vision, machine learning, and natural language processing to promote variety), and stratified sampling~(Al-Zaidy \etal \cite{al2017machine} only sampled one chart per PDF file based on their observation that most charts from the same document have the same layout which hurts diversity).

For computer-aided generated chart corpora described in Section~\ref{computerAidedGeneration},  diversity needs to be enforced during the computer-generating process.
There are two ways to increase diversity in this setting: diversifying underlying datasets and diversifying style parameters.

\bpstart{Diversify underlying datasets.} As stated in Section~\ref{computerAidedGeneration}, people randomize data types and tune the underlying distributions of data attributes to generate a variety of synthetic datasets for plotting. It has also been observed that people enumerate combinations of data attributes to increase the number of plottable charts and diversify styles; e.g., Ma \etal~\cite{ma2018scatternet} used 757 datasets from the PyDataset library~\cite{Pydataset}, and combined all possible pairs of columns in each of them to form a scatterplot, resulting in 50677 different scatterplots.
To obtain diverse public datasets, people look for those with various topics, years, and cultures. For example, Methani~\etal~\cite{methani2020plotqa} considered online data sources such as World Bank Open Data~\cite{Wbod} and Global Terrorism Database~\cite{Gtd} which contain statistical data about a variety of factors including fertility rate and coal production across years and countries. 

\bpstart{Diversify style parameters.} Another typical and important practice in creating computer-generated chart corpora is diversifying style parameters in the codes. We summarize commonly considered style parameters in Table~\ref{tbl:stylePara}, including color, presence of grid lines, legend and axis location, legend and mark orientation, title location, and font size and family. Some other less frequently used parameters include label orientation~\cite{kafle2018dvqa}, tick size and orientation~\cite{zhou2021reverse}, etc. 

\begin{table*}[ht!]
    \centering
    \caption{Commonly-seen style parameters that can be randomized to increase the diversity of a computer-generated chart corpus.}
    \begin{tabular}{ll}
    \toprule
     Style Parameter    &   Relevant Corpora\\
     \midrule
      fill color                 & \cite{sharma2019chartnet,chagas2018evaluation,kahou2017figureqa,chaudhry2020leaf,methani2020plotqa,zhou2021reverse,cliche2017scatteract
      }  \\
presence of grid lines               & \cite{kafle2018dvqa,chagas2018evaluation,kahou2017figureqa,chaudhry2020leaf,methani2020plotqa,poco2017reverse,cliche2017scatteract,chen2019towards}  \\
legend location & \cite{kafle2018dvqa,kahou2017figureqa,chaudhry2020leaf,methani2020plotqa,zhou2021reverse,poco2017reverse} \\
axis location & \cite{poco2017reverse, cliche2017scatteract} \\
legend orientation        & \cite{kafle2018dvqa,kahou2017figureqa}   \\
mark orientation                 & \cite{kafle2018dvqa,zhou2021reverse,chen2019towards}  \\
font family and size & \cite{chaudhry2020leaf,methani2020plotqa,zhou2021reverse,poco2017reverse,chen2019towards} \\
title location & \cite{chaudhry2020leaf,zhou2021reverse,cliche2017scatteract} \\
     \bottomrule
    \end{tabular}
\label{tbl:stylePara}
\end{table*}

\bpstart{Diversify visual questions and captions.}
As we mentioned in Section~\ref{sec:annotationMethod}, template-based QA or QC annotation generally is limited by poor linguistic variations, which hurts the generalizability of developed models to real-world human-raised questions. Thus, new techniques have been devised recently to alleviate this problem. For example, Chen~\etal~\cite{chen2019figure} designed a large number of templates to achieve more than 200 possible variations with the same meaning for high-level captions; similarly, Singh and Shekhar~\cite{singh2020stl} manually created 3 to 8 paraphrase variations of question templates to boost the diversity and naturalness of the questions significantly. Methani~\etal~\cite{methani2020plotqa} took a different approach to diversify QA templates: they first gathered a larger set of annotators to create questions based on 1400 charts, then manually analyzed the questions collected and synthesized 74 question templates, and finally hired in-house annotators to manually paraphrase the templates carefully to avoid unnaturalness; the final PlotQA dataset is much closer to the real-world challenge of reasoning over charts. More recently, Masry~\etal~\cite{masry2022chartqa} abandoned templates and adopted T5~\cite{raffel2020exploring}, a large-scale language model that is trained on very large public data and was shown to learn general linguistic properties and variations~\cite{brown2020language}, to generate human-like questions with adequate lexical and syntactic variations automatically.

%% file: sections/9_challenges_and_opportunities.tex
\section{Challenges and Opportunities}
Section \ref{sec:reuse_corpus} mentions that very few corpora are reused or transformed in subsequent works. As a result, it is difficult to evaluate and compare related analysis techniques and measure research progress. This leads to the question of whether 
we can build benchmark corpora for the same chart analysis tasks. In addition, are there gaps in the current literature that entail the need to create new types of corpora with new types of annotations? 
In this section, we reflect on these questions based on our survey in the previous 6 sections. We first identify under-explored problems and research opportunities in corpora-based automated chart analysis, then suggest approaches for building benchmark chart corpora to support these research efforts.

\subsection{Under-explored Problems and Research Opportunities}\label{sec:9.1}
\begin{figure}[ht]
    \centering
    \includegraphics[width=0.475\textwidth]{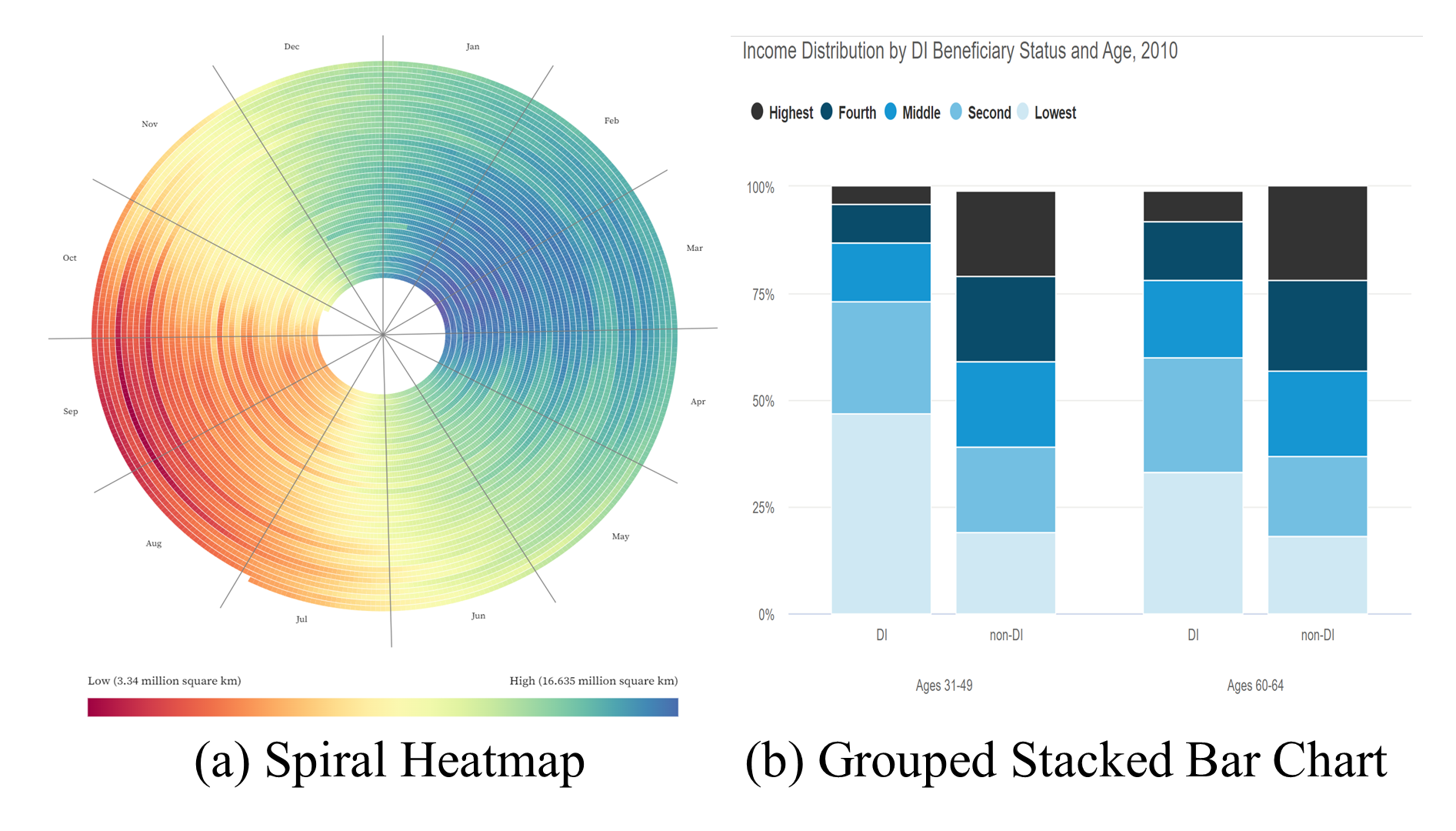}
    \caption{A real-world chart design can be classified into more than one chart type. (a) from~\cite{SpiralHeatmap}: both a spiral plot and a heatmap; (b) from~\cite{GroupStackedBar}: both a grouped bar chart and a stacked bar chart.
    }
    \label{fig:multiTagExample}
    \end{figure}

\bpstart{RO1: Beyond Chart Types.}
For semantics extraction tasks, chart type is often used as a high-level description to classify input charts. 
Automatically categorizing chart types can support downstream applications such as redesign and reuse~\cite{savva2011revision,battle2018beagle,deng2022visimages}. However, chart typologies used in the papers are sometimes not consistent. For example, histogram was listed as one of the chart types considered in Saleh \etal\cite{saleh2015learning}, while it was subsumed under the bar chart type in Li \etal~\cite{li2022structure}. Also, the descriptive power of the concept of a chart type has its limitations. First, it is possible that a chart design may be classified into more than one chart type; for example, Figure~\ref{fig:multiTagExample}(a) is both a spiral plot (focusing the layout) and a heatmap (focusing the colors), and Figure~\ref{fig:multiTagExample}(b) can be described as a grouped bar chart or a stacked bar chart. In such cases, it is hard to decide on just one category.
Second, within one chart type, many design variations can exist. It is important for researchers to specify the exact scope of their work; for example, Chen~\etal~\cite{chen2019figure} explicitly specified horizontal or vertical single bar charts as their scope, and Chaudhry~\etal~\cite{chaudhry2020leaf} mentioned stacked/grouped/single bar charts as their scope. Many other papers~\cite{al2017machine,luo2021hybrid}, however, defined and used chart types casually, making it hard to measure a model's exact capacity. 
The limited descriptive power and the lack of consistently defined chart typologies may have contributed to the difficulties of building benchmark corpora. 



Instead of high-level chart typologies, feature tags can be a better way to describe a given chart --- multiple tags can be used together to specify the mark types and visual designs. Taking Figure~\ref{fig:multiTagExample}(b) as an example, tags we can assign to it include ``bar'', ``stacked'', and ``grouped'', which span a more complete description compared to a vague chart category.
We can further add ``grid'' to indicate the layout information, which in most cases is not included in chart typologies. Thus, one can use different levels of tags to represent multi-level semantics information presented in a chart, which is potentially more helpful for chart analysis tasks.

Taking a step further, we can think about how modern charting tools generate visualizations: they have long moved beyond chart types to adopt the Grammar of Graphics~(GoG) paradigm~\cite{wilkinson2012grammar}; examples include Vega-Lite \cite{satyanarayan2016vega} that provides declarative specifications of grammar primitives,  Charticulator~\cite{ren2018charticulator} that proposed a constraint-based chart layout framework to achieve bespoke designs, and Data Illustrator~\cite{liu2018data} that incorporated data binding into direct manipulation of chart elements. It is worth considering how we can 
label and decompose charts into graphical primitives. 
Some works have started to develop techniques sharing this philosophy, e.g., the graphical element update taxonomy proposed in Chartreuse~\cite{cui2021mixed}. More research needs to be investigated to make it generalizable to a broader range of charts and tasks.



\bpstart{RO2: Beyond Chart Similarity.}
Some works, 
e.g., Hu~\etal~\cite{hu2019vizml}, Dibia and Demiralp~\cite{dibia2019data2vis}, and Li~\etal~\cite{li2022structure}, 
 recommend or retrieve charts based on similarities in visual structure or style. 
Although similarity is an informative metric in many situations, some applications can require other types of derived chart properties. 
For example, when visualization creators are seeking design ideas, similarity may not be their primary desired criterion; instead, they prefer alternative or bespoke designs to broaden the scope of consideration \cite{bako2022understanding}. 

Chart quality is another under-explored derived property. In our surveyed papers, only Fu~\etal~\cite{fu2019visualization} focused on chart quality, trying to rank charts regarding aesthetics or memorability scores automatically.
Apart from aesthetics, we would like to point out that chart quality has other dimensions, such as effectiveness, i.e., to what extent a chart design is suitable for visualizing given datasets. 
In general, the automatic assessment of chart quality~(including aesthetics and suitableness) remains an open research question.

\bpstart{RO3: Tool/Source-Agnostic Chart Analysis.}
Many works use a chart corpus with a narrow scope and low diversity, assuming that the charts belong to a specific type, created by specific tools or from specific sources. 
These works thus have listed generalizability as one of their limitations and acknowledge tool/source-agnostic chart analysis techniques or systems as an important problem to be addressed in future work. For example, some systems such as Qian~\etal~\cite{qian2020retrieve} and Cui~\etal~\cite{cui2019text} rely on predefined templates to generate the final infographics, thus are not expected to scale well in terms of design variation~\cite{cui2021mixed}; Obeid and Hoque~\cite{obeid2020chart} would like to create larger corpora that cover more diverse domains further to improve the generalizability of their chart summarizing model; the D3 search engine~\cite{hoque2019searching} includes supporting databases containing diverse chart collections, helping to discover differences regarding design patterns across a variety of sources in their future work; the chart retrieval technique in Li~\etal~\cite{li2022structure}, which is built solely on charts form Plotly~\cite{Plotly} and required consistent usage and grouping of SVG elements, could fail to function well with charts from other tools like D3~\cite{bostock2011d3} and Data Illustrator~\cite{liu2018data}. Thus, increasing the diversity of chart corpora to create tool/source-agnostic techniques for automated chart analysis is a consensus within this field, and remains a significant research problem.

\bpstart{RO4: Design Generation with More Diverse Corpus.} Current research on automatic generation of chart design mostly relies on a chart corpus in the program format from a single charting tool (e.g., Zhao~\etal~\cite{zhao2020chartseer} and Dibia and Demiralp~\cite{dibia2019data2vis}). This practice limits corpus diversity, which in turn potentially hurts the diversity of generated designs. Ample research opportunities are available when we include corpora in SVG or raster image formats as the basis for chart generation, which require novel techniques to synthesize visual structures and designs from various sources. 

\bpstart{RO5: Systematic Methods to Measure Diversity.} \revise{As discussed in Section~\ref{sec:diversity}, researchers are using a variety of empirical methods to enhance diversity in chart corpora. However, 
to date, there have been no metrics to quantify or measure diversity. 
There is a clear need for systematic methods to evaluate chart diversity within a corpus and compare diversity between corpora. Such methods can guide chart selection processes and enhance the rigor of automated chart analysis research. 
}

\bpstart{RO6: Interactive and Animated Charts.}
Unlike online bitmap charts~(.png, .jpeg), SVG-based charts are
oftentimes interactive and animated~\cite{battle2018beagle}. However, the logic for interactions and animations is currently specified using JavaScript in most cases rather than being part of the SVG specification that is extractable. In all the SVG-based corpora we surveyed, none has annotations regarding interactive or animated behaviors. The D3 search engine \cite{hoque2019searching}, for instance, discusses interaction support as a limitation and future work. How to automatically capture, represent, understand, and extract interactivity and animation remains under-explored and would potentially facilitate new research ideas.

\subsection{Desired Properties of Benchmark Corpora}

With the goal of building benchmark corpora and the open opportunities described in Section~\ref{sec:9.1} in mind, we discuss the desired properties~(DP) of benchmark corpora below.
\bpstart{DP1: Enhance Chart Diversity within a Corpus.}
Diversity in terms of chart source, employed charting tool, chart design, and visual style plays a vital role in the generalizability and robustness of chart analysis techniques. The standard practices 
presented in Section~\ref{sec:diversity} can be applied to achieve this desired property. 

In addition, we have found little effort in current practices to enhance diversity in terms of chart format. As we discussed in Section~\ref{sec:format}, both the bitmap and vector graphics formats have their unique pros and cons, e.g., parsing a vector-format chart might give more accurate results compared to extracting the same information from a bitmap image, while collecting qualified vector graphics might be more laborious and error-prone. In this sense, it is ideal that a benchmark corpus can include both formats for each chart and maintain a set of annotations (\eg bounding boxes and mark types) that are applicable to both formats. Charts in the program format can also be considered, while it would be more difficult than merging bitmaps with vector graphics because program-format charts are text-based and their grammar varies according to the underlying languages.

A diverse corpus can provide a strong foundation for research investigations in \textbf{RO2} (Beyond Chart Similarity), \textbf{RO3} (Tool/Source-Agnostic Chart Analysis), \textbf{RO4} (Design Generation with More Diverse Corpus), \revise{and \textbf{RO5} (Quantifying Diversity)}.

\bpstart{DP2: Multi-level Fine-Grained Annotations.}  
Most of the existing annotations are either at the chart level~(e.g. chart type, source data) or about position information~(e.g., mark bounding boxes). Only a few corpora contain finer-grained annotations at the component level like encodings and layout which require deeper extraction of semantics, and these corpora usually have a narrow scope and limited diversity. The lack of fine-grained annotations makes it difficult to reuse some large-scale and diverse corpora 
for new tasks. We expect a benchmark dataset to contain multi-level fine-grained annotations, which can support a variety of chart analysis tasks and may lead to novel research questions and application scenarios. Apart from the commonly seen annotation types described in Section~\ref{sec:annotationType}, chart feature tags and effectiveness labels 
should also be considered. With a set of multi-level fine-grained annotations, it is easier to transform a benchmark corpus into a desired one by filtering specific annotation values; e.g., one can set the chart tags to the combination of ``grouped'' and ``bar'' to receive grouped bar charts in the corpus. With such annotations consistently applied across different collecting sources~(\textbf{DP1}), it is also more straightforward to test the generalizability of developed techniques.

The task of generating natural language descriptions can also benefit from fine-grained annotations that establish correspondences between chart components such as marks and encodings with text elements such as tokens, phrases, and sentences~\cite{kong2014extracting}. Such annotations can also enable automatic synchronization between chart and text components for applications beyond synthesizing descriptions, e.g., dynamic presentation of relevant charts for enhanced reading~\cite{badam2018elastic}, and automatic linking text and chart elements in dynamic layouts~\cite{sultanum2021leveraging}. This thread of work is not included in this report because (1) their corpora mainly consist of storytelling articles and papers~\cite{latif2021kori} whose property space could be different, and (2) the consideration of document layout that is beyond the charts themselves~\cite{sultanum2021leveraging}. 
Thus, we leave a more detailed analysis of these techniques as future work.

It is worth mentioning that there have been efforts to improve the quality of annotations in benchmark chart corpora. One example is the ICPR CHART-Infographics dataset~\cite{ICDAR} used in the CDAR competition on raw data extraction and visual question answering. Each chart in the dataset has a corresponding JSON representation of annotations over chart type, text values and roles, axes and legend, underlying raw data, and QA pairs.
Their annotation tool~\cite{AnnotationTool4ICDAR} has also been released. This competition highlights the importance of benchmark corpora for developing and comparing models, and the community's awareness of the need for such corpora with high-quality annotations. Still, further work is needed to create fine-grained corpora for different tasks and use cases.

A corpus with aforementioned multi-level annotations can support research efforts in \textbf{RO1} (Beyond Chart Types), \textbf{RO2} (Beyond Chart Similarity), \textbf{RO4} (Design Generation with More Diverse Corpus), \revise{and \textbf{RO5} (Quantifying Diversity)}.

\bpstart{DP3: Interactivity and Animation Understanding.}
Ideally, a benchmark corpus also contains meta-information or annotations about the interactive or animated behaviors in SVG charts. Two aspects of obtaining such data shall be researched: (1) the semantic abstractions or tags for describing interactivity and animation, and (2) the methods for capturing and understanding interactivity and animation. Some previous works have made efforts in these two aspects: Park~\etal~\cite{park2008designers} and Myers~\etal~\cite{myers2008designers} examined how designers design and describe interactive behaviors, and Raji~\etal~\cite{raji2020dataless} developed a system called Loom to capture and share interactive visualization in the Tableau application. More research needs to be investigated to help record the interactivity and animation information in a corpus, which can further support research efforts in \textbf{RO1} (Beyond Chart Types) and \textbf{RO6} (Interactive and Animated Charts).

\subsection{Desired Tools for Creating Benchmark Corpora}
In this section, we propose multiple desired tools~(DT) that can facilitate the creation of benchmark corpora with properties described in \textbf{DP1}-\textbf{DP3}.


\bpstart{DT1: Smart web crawler.} Given \textbf{DP1} and \textbf{DP3}, we may need to collect interactive SVG images from a variety of sources where the composition structure of SVG charts on the web differ. There are some special cases where the web crawler needs to take additional care to collect a complete SVG file. For example, in the Plotly Gallery of bar charts\footnote{https://plotly.com/python/bar-charts/}, there can be multiple SVG elements in the HTML for a single interactive SVG chart, one is for rendering the chart and the others are for interaction controls; also, the legend for an SVG chart is usually stored in a separate SVG element in the HTML.
Thus, the following features would be useful: (1) automatic merging and filtering of SVG chart elements, (2) identifying and recording SVG elements for interaction and animation controls, and (3) necessary post-processing such as crop and resize for both vector graphics and bitmaps.
Implementing such a smart web crawler that works for both interactive SVG charts and bitmaps is important to the corpus quality;
otherwise, intensive labor would be expected to ensure the quality of SVG charts.

\bpstart{DT2: Tools to pre-process and clean up SVG.} As discussed in Section~\ref{sec:format}, the embedded semantic information in SVG charts from the wild is not always accurate or reliable. Apart from the web crawler, we need tools to accurately extract SVG elements and unify them in a consistent format to achieve \textbf{DP1}, \textbf{DP3} and prepare for \textbf{DP2}. The features of such tools include but are not limited to:
\begin{enumerate}
    \item Analyzing a given \textit{<path>} element to identify its mark type~(rectangle, circle, line, polygon, etc.).
    \item Filtering graphical elements that are not part of the chart semantics, such as watermarks and backgrounds.
    \item Merging separate text elements that belong to the same label or title together.
    \item Recording accurately the positions and the visual style attributes of every SVG element.
\end{enumerate}

\bpstart{DT3: Mix-initiative annotating system.}
 Obtaining high-quality annotations is generally expensive and time-consuming, especially for complex annotations~(\textbf{DP2}) requiring careful examination over charts. To this end, dedicated research in human-AI collaboration for the annotation process is necessary. To achieve an effective mix-initiative labeling system, we need to consider for each annotation type, what steps can be automated by the computers and what steps shall be operated or examined by humans. Let's take QA pair annotations as an example. Most current practices for annotating QA pairs are either through templates~\cite{chen2019figure,singh2020stl} or employing annotators~\cite{methani2020plotqa,kim2020answering}. Given the recent advances in large language models~\cite{floridi2020gpt}, it is possible that the annotation system first generates a set of questions and lets the annotator identify valid ones. After that, the system can utilize state-of-the-art QA techniques to propose answers to selected questions, whose tokens can be interactively edited by the annotator. 
 During this collaboration, additional finer-grained annotations, such as references between texts~(questions or answers) and chart components~\cite{kong2014extracting} can be further introduced, adding more details~(\textbf{DP2}). This kind of human-in-the-loop annotating process is expected to significantly reduce the cost of obtaining annotations and boost the research in automated chart analysis.

%% file: sections/10_conclusion.tex
\section{Conclusion}
In this state-of-the-art report, we review 56 chart corpora created or used for automated chart analysis. Our analysis is based on a three-level task taxonomy (goal, method, output) and five corpus properties (format, scope, collection method, annotations, and diversity). We argue that there is a need to create benchmark corpora of higher diversity with multi-level finer-grained annotations for various chart analysis tasks. We identify new research opportunities in building tools for creating benchmark corpora, and discuss how such corpora can form the foundation for future advances in automated chart analysis research. 

%% file: sections/ShortBio.tex
\section*{Short Biographies of Authors}

\subsection*{Chen Chen}
Chen Chen is a PhD student working with Professor Zhicheng Liu in
the Human-Data Interaction Research Group affiliated with the Human-Computer Interaction Lab (HCIL) at University of Maryland, College Park. His research focuses on data visualization grammar, visualization understanding and reuse, and human-centered AI.

\subsection*{Zhicheng Liu}
Zhicheng Liu is an assistant professor of computer science at the University of Maryland College Park (UMD). Before joining UMD, he was a research scientist at Adobe Inc. He directs the Human-Data Interaction research group, which is affiliated with the Human-Computer Interaction Lab at UMD. His past and current research focuses on data visualization grammar and frameworks \cite{liu2021atlas,liu2011network,thompson2020understanding}, visualization design and authoring tools \cite{liu2018data,thompson2021data,kim2016data,liu2014ploceus}, and techniques to analyze, deconstruct, and reuse visualizations  \cite{saleh2015learning}.